\documentclass[sigplan, screen, noacm]{acmart}

\setcopyright{rightsretained}
\acmDOI{}

\acmConference[]{February 2024}{Zurich}{Switzerland}
\acmISBN{}

\usepackage{adjustbox}
\usepackage{ifthen}
\usepackage{caption}
\usepackage{subcaption}
\usepackage{xfp}
\usepackage{calc}
\usepackage{array}
\usepackage{xcolor}
\usepackage{booktabs}
\usepackage{multirow}
\usepackage{longtable}
\usepackage{bbm}
\usepackage{tikz}
\usepackage{pgfplots}
\usetikzlibrary{positioning}
\usetikzlibrary{calc}
\usetikzlibrary{backgrounds} 
\pgfplotsset{compat=1.16}
\usepackage{listings}
\usepackage{color}
\usepackage[noend]{algorithmic}
\usepackage{algorithm}
\usepackage{float}
\usepackage{bm}
\usepackage{eqnarray}
\usepackage{amsmath}
\usepackage{enumitem}
\usepackage{multicol}
\usepackage{supertabular}

\newcommand{\ignore}[1]{}

\newcommand{\tool}{\textsc{DeepCode AI Fix}}
\newcommand{\github}{GitHub}
\newcommand{\codereduce}{\textsc{CodeReduce}}

\newcommand{\mergeback}{\textsc{MergeBack}}

\newcommand{\ddmin}{\textsc{DDMin}}

\newcommand{\PromptSource}{${\color{red}\texttt{CODE}^q_\texttt{PRE}}$}

\newcommand{\FewShotSource}{${\color{red}\texttt{CODE}^i_\texttt{PRE}}$}
\newcommand{\FewShotTarget}{${\color{teal}\texttt{CODE}^i_\texttt{POST}}$}
\newcommand{\FewShotSourceOne}{${\color{red}\texttt{CODE}^1_\texttt{PRE}}$}
\newcommand{\FewShotTargetOne}{${\color{teal}\texttt{CODE}^1_\texttt{POST}}$}
\newcommand{\FewShotSourceLast}{${\color{red}\texttt{CODE}^f_\texttt{PRE}}$}
\newcommand{\FewShotTargetLast}{${\color{teal}\texttt{CODE}^f_\texttt{POST}}$}
\newcommand{\PromptRule}{${\color{purple}\texttt{RULE}}$}
\newcommand{\PromptMessage}{${\color{blue}\texttt{DESCRIPTION}}$}
\newcommand{\PromptSystem}{${\color{blue}\texttt{SYSTEM:}}$}
\newcommand{\PromptUser}{${\color{cyan}\texttt{USER:}}$}
\newcommand{\PromptAssistant}{${\color{green}\texttt{ASSISTANT:}}$}

\newcommand{\AST}{\textsc{AST}}
\newcommand{\Local}{\textsc{Local}}
\newcommand{\FileWide}{\textsc{FileWide}}
\newcommand{\SecurityLocal}{\textsc{SecurityLocal}}
\newcommand{\SecurityFlow}{\textsc{SecurityFlow}}

\newcommand{\ExactMatchAtKMetric}{\textsc{ExactMatch@$k$}}

\newcommand{\PassAtKMetric}{\textsc{Pass@$k$}}

\newcommand{\DoesFix}{\ensuremath{\mathrm{DoesFix}}}
\newcommand{\NoNewIssues}{\ensuremath{\mathrm{NoNewIssues}}}

\newcommand{\NumRules}{$156$}

\newcommand{\NumTestSamples}{$1'818$}

\newcommand{\JsWindowThreeData}{{\textsc{Window@3}}}
\newcommand{\JsWindowFiftyData}{{\textsc{LongContext}}}
\newcommand{\JsReducedData}{{\textsc{CodeReduced}}}
\newcommand{\JsOriginalData}{{\textsc{FullOriginal}}}

\newcommand{\algorithmicinput}{\textbf{Input}}
\newcommand{\algorithmicoutput}{\textbf{Output}}
\newcommand{\INPUT}{\item[\algorithmicinput]}
\newcommand{\OUTPUT}{\item[\algorithmicoutput]}

\newcommand{\replacementmapping}{\ensuremath{r_{c\to p}}}
\newcommand{\linemapping}{\ensuremath{l_{c\to C}}}

\newcommand{\fixedlines}{\ensuremath{\mathcal{L}_{\mathrm{fixed}}}}
\newcommand{\sourcelines}{\ensuremath{\mathit{Lines}_C}}
\newcommand{\plines}{\ensuremath{\mathit{Lines}_p}}

\newcommand{\gitdiff}{\texttt{gitdiff}}

\pgfdeclarelayer{foreground}
\pgfsetlayers{main,foreground}

\newcommand\realnumberstyle[1]{}
\makeatletter
\newcommand{\linecolors}[4]{
	{\realnumberstyle{#1}}
	\begingroup
	\lst@basicstyle
	\ifnum\value{lstnumber}<#2
		\color{white}
	\else
		\ifnum\value{lstnumber}>#3
			\color{white}
		\else
			\color{#1}
		\fi
	\fi
	\rlap{\hspace*{\lst@numbersep}
		\color@block{\linewidth}{\ht\strutbox}{\dp\strutbox}
	}
	\endgroup
}
\makeatother

\makeatletter
\newcommand{\HL}[1]{%
	{\realnumberstyle{}}%
	\begingroup%
	\lst@basicstyle%
	\color{#1}%
	\rlap{%
		\color@block{\linewidth}{\ht\strutbox}{\dp\strutbox}%
	}%
	\endgroup%
}%
\makeatother

\DeclareRobustCommand{\RoundedFrameWord}[1]{%
	\tikz[overlay]\node [draw=red,anchor=text,inner xsep=1pt,inner ysep=1pt, fill=white,rectangle,rounded corners=2pt] {#1};\phantom{#1}
}

\makeatletter
\newcommand{\RoundedFrameLine}{%
	{\realnumberstyle{}}%
	\begingroup%
	\lst@basicstyle%
	\begin{tikzpicture}[overlay]%
		\begin{pgfonlayer}{foreground}%
			\node [draw=red,anchor=text,inner xsep=1pt,inner ysep=1pt, fill=white,rectangle,rounded corners=2pt, fill opacity=0.0] {\makebox[\linewidth]{\rule[-.31\baselineskip]{0pt}{1.02\baselineskip}}};%
		\end{pgfonlayer}%
	\end{tikzpicture}%
	\endgroup%
}%
\newcommand{\RoundedFrameThreeLine}{%
	{\realnumberstyle{}}%
	\begingroup%
	\lst@basicstyle%
	\tikz[overlay]\node [draw=red,anchor=text,inner xsep=1pt,inner ysep=1pt, fill=white,rectangle,rounded corners=2pt, fill opacity=0.0] {\makebox[\linewidth]{\rule[-3.1\baselineskip]{0pt}{3.8\baselineskip}}};%
	\endgroup%
}%
\makeatother

\definecolor{myblue}{rgb}{0.6, 0.85, 1}
\definecolor{mypurple}{rgb}{0.77, 0.77, 1}
\definecolor{myred}{rgb}{1,0.8,0.8}
\definecolor{myyellow}{rgb}{0.94,0.9,0.66}
\definecolor{mygreen}{rgb}{0.75, 0.95, 0.75}
\definecolor{myorange}{rgb}{1, 0.85, 0.73}
\definecolor{mylinegray}{gray}{0.5}
\definecolor{mydrawgray}{gray}{0.7}
\definecolor{myemptygray}{gray}{0.8}
\definecolor{mybrown}{rgb}{0.80, 0.52, 0.28}

\definecolor{commentsColor}{rgb}{0.497495, 0.497587, 0.497464}
\definecolor{keywordsColor}{rgb}{0.000000, 0.000000, 0.635294}
\definecolor{stringColor}{rgb}{0.558215, 0.000000, 0.135316}

\lstdefinelanguage{js}{
	basicstyle=\ttfamily\scriptsize\linespread{1.04},
	keywords={var, for, in, if, continue, constructor, this, super, case, typeof, instanceof, new, const, let, NaN, console, return, throw, import, catch, set, get, typeof, new, true, false, catch, function, return, null, catch, switch,  while, do, else, case, break},
	ndkeywords={class, export, boolean, throw, implements, import, this},
}

\lstdefinelanguage{jssmall}{
	basicstyle=\fontsize{6.5}{8}\selectfont\ttfamily\linespread{1.04},
	keywords={var, for, in, if, continue, constructor, this, super, case, typeof, instanceof, new, const, let, NaN, console, return, throw, import, catch, set, get, typeof, new, true, false, catch, function, return, null, catch, switch,  while, do, else, case, break},
	ndkeywords={class, export, boolean, throw, implements, import, this},
}

\lstdefinelanguage{jsappdx}{
	basicstyle=\fontsize{5}{6}\selectfont\ttfamily\linespread{1.04},
	keywords={var, for, in, if, continue, constructor, this, super, case, typeof, instanceof, new, const, let, NaN, console, return, throw, import, catch, set, get, typeof, new, true, false, catch, function, return, null, catch, switch,  while, do, else, case, break},
	ndkeywords={class, export, boolean, throw, implements, import, this},
}

\DeclareRobustCommand{\TightColoredWord}[2]{{\setlength{\fboxsep}{0pt}\colorbox{#1}{#2\vphantom{gd}}}}
\DeclareRobustCommand{\LessTightColoredWord}[2]{{\setlength{\fboxsep}{1pt}\colorbox{#1}{#2\vphantom{gd}}}}

\newcommand{\mulro}[3][0]{%
	\multirow{#2}{*}[-\fpeval{#1*#2}pt]{#3}
}

\begin{document}

\title[DeepCode AI Fix]{DeepCode AI Fix: Fixing Security Vulnerabilities with Large Language Models}
%

\author{Berkay Berabi}
\email{berkay.berabi@snyk.io}
\affiliation{%
	\institution{Snyk}
	\city{Zurich}
	\country{Switzerland}
}
\author{Alexey Gronskiy}
\email{alex.gronskiy@snyk.io}
\affiliation{
	\institution{Snyk}
	\city{Zurich}
	\country{Switzerland}
}

\author{Veselin Raychev}
\email{veselin.raychev@insait.ai}
\affiliation{
	\institution{INSAIT, Sofia University}
	\city{Sofia}
	\country{Bulgaria}
}

\author{Gishor Sivanrupan}
\email{sgishor@gmail.com}
\affiliation{
	\institution{}
	\city{Zurich}
	\country{Switzerland}
}

\author{Victor Chibotaru}
\email{viktor.chibotaru@snyk.io}
\affiliation{
	\institution{Snyk}
	\city{Zurich}
	\country{Switzerland}
}

\author{Martin Vechev}
\email{martin.vechev@inf.ethz.ch}
\affiliation{
	\institution{ETH Zurich}
	\city{Zurich}
	\country{Switzerland}
}
\renewcommand{\shortauthors}{Berabi, et al.}


%

%
\begin{abstract}
	The field of automated program repair has attracted substantial interest over the years, but despite significant research efforts, creating a system that works well for complex semantic bugs such as security vulnerabilities has proven difficult. A promising direction to addressing this challenge is by leveraging large language models (LLMs), which are increasingly used to solve various programming tasks such as code generation.

	In this paper, we investigate the effectiveness of LLMs for solving such code-repair tasks. We show that the task is difficult as it requires the model to learn long-distance code relationships, a task that inherently relies on extensive amounts of training data. At the same time, creating a large, clean training or evaluation datasets for complex program bugs and their corresponding fixes is costly and non-trivial. We propose a technique to address these challenges with a new approach for querying and fine-tuning LLMs. The key idea is to leverage program analysis to limit the LLM’s attention mechanism on the portions of code needed to perform the fix, drastically reducing the amount of required training data. Concretely, for both training and inference, rather than feeding the entire program to the LLM, we reduce its code to a much shorter snippet that contains the reported defect together with the necessary context – and use that instead.

	Our evaluation shows that this code reduction approach substantially improves both readily available models such as GPT-4 using few-shot learning, as well as fine-tuning models. To train and evaluate our system, we also created a new and comprehensive code fixing dataset by extensively labeling 156 non-trivial bug patterns (including 40 security rules). These patterns are non-trivial and require complex interprocedural dataflow to discover. Our best resulting system based on Mixtral-8x7B can remove more than 80\% of the reported defects while exactly matching the human fix in between 10 and 50\% of cases, outperforming baselines based on GPT-3.5 and GPT-4, or based on window-based models such as TFix.
\end{abstract}

\maketitle

\section{Introduction}

The rapid increase in the number of software systems has led to an increased demand for methods and tools that can detect potential defects in these systems. Indeed, existing testing and bug-finding tools, ranging from test generation to dynamic and static analysis, often return thousands of high-quality findings~\cite{coverity}. However, due to the sheer volume of discovered defects, most of the reports are not addressed by developers. Despite various efforts aiming to improve the development process and improve the situation~\cite{googleStaticAnalysis}, the most promising long-term path to addressing this issue in a systematic manner remains automatic bug fixing.

Prior to using large language models (LLMs), automated fixing tools mostly focused on a small set of bugs~\cite{sapfix}, trivial few-line changes~\cite{hoppity, SemGrepFix} or formatting issues~\cite{sapfix, ESLintRules}. As LLMs were created, it was shown that they can also be useful in fixing lint-like issues~\cite{TFix, RING}. With even larger LLMs, a number of software security providers announced features that mostly use OpenAI GPT models~\cite{gpt4, gpt35} in an attempt to automatically fixing more complex security issues discovered by static analyzers~\cite{CodeQLAutofix, SemGrepAutofix}. Other security providers announced automatic fixing tools based on custom trained models~\cite{SnykCodeAutofix, VeracodeAutofix}. However, outside of marketing materials that display the user interface of these tools, there is little understanding of the underlying accuracy of the provided fixes. One of our contributions in this work is to address this issue by designing a dataset of vulnerabilities and their fixes, using which one can now evaluate the capabilities of existing and as well as any models proposed in future.


\textbf{A dataset of security and semantic code fixes.}
To address the lack of evaluation data, we performed data collection and labeled thousands of commits collected from open source repositories. Using $380'000$ candidate fix commits and tens of labelers with coding and security background, we assembled a dataset of over $5'000$ manually labelled fix examples. We then split this data according to the licenses of the code -- permissively licensed data can be used for training or fine-tuning, and other data used only for evaluation. Using this method, we could protect against train/test leakage in evaluations as some of the models we use (like StarCoder~\cite{starcoder}) are trained only on permissively licensed data.
Our dataset is based on code scanning results of Snyk Code SAST engine~\cite{SnykCode}. Snyk Code operates on source code and is one of the fastest engines~\cite{SnykCodeSpeed} with an extensive set of security rules. Each fix in our dataset consists of code where there is an alarm raised by Snyk Code and code where the alarm is not raised. In addition, human labelers agreed that the change correctly fixes the given alarm. Thus, our dataset has its result validated by a human that confirmed the change was not due to false positives or false negatives of the engine.
\paragraph{Key challenge: difficult to obtain training data}
A core challenge when using machine learning for program repair is that it is inherently difficult to obtain a sufficiently large, clean dataset consisting of pairs of buggy and fixed code versions for a given program issue. While the history of many open source projects is available on \github{}, obtaining a dataset of fixes has so far only been done successfully for simple few-line local static analyses~\cite{TFix, RING}. The reason underlying this difficulty is that with deeper semantic issues, it is possible to observe program changes where a defect report is present in one version and absent in another, yet the core program error is not fixed. For example, the change may turn a problematic branch into dead code, some unrelated change may accidentally make a security vulnerability in the code exploitable, a code change may modify many more places than the bug or the change may affect the approximation of the static analyzer without affecting the actual presence of the reported issue. We investigate the frequency of these problems in Section~\ref{sec:data}.

\begin{figure*}
	\hspace{2cm}
	\small
	\begin{tikzpicture}

		\node at (-0.2, 1.8) {\scriptsize Training data (git commits)};

		\begin{scope}[shift={(-1,1.5)}]
			\draw[thick] (0,0) -- (0.6,0);
			\draw[thick] (1,0) -- (1.6,0);
			\draw[very thick,color=red] (-.05,-0.05) -- (0.6,-0.05);
			\draw[thick] (0,-0.1) -- (0.5,-0.1);
			\draw[thick] (1,-0.1) -- (1.5,-0.1);
			\draw[very thick,color=green] (0.95,-0.15) -- (1.7,-0.15);
			\draw[thick] (0,-0.2) -- (0.7,-0.2);
			\draw[thick] (1,-0.2) -- (1.7,-0.2);
			\draw[thick] (0,-0.3) -- (0.6,-0.3);
			\draw[thick] (1,-0.3) -- (1.6,-0.3);
			\draw[very thick,color=red] (-.05,-0.35) -- (0.8,-0.35);
			\draw[very thick,color=green] (0.95,-0.35) -- (1.8,-0.35);
			\draw[thick] (0,-0.4) -- (0.6,-0.4);
			\draw[thick] (1,-0.4) -- (1.6,-0.4);
			\draw[thick] (0,-0.5) -- (0.6,-0.5);
			\draw[thick] (1,-0.5) -- (1.6,-0.5);
			\draw[very thick,color=green] (0.95,-0.55) -- (1.8,-0.55);
			\draw[thick] (1,-0.6) -- (1.6,-0.6);
			\draw[very thick,color=green] (0.95,-0.65) -- (1.7,-0.65);
			\draw[thick] (1,-0.7) -- (1.6,-0.7);
		\end{scope}

		\node[scale=0.05] (CodeReduceTrain) at (3.75,1.1) {\includegraphics{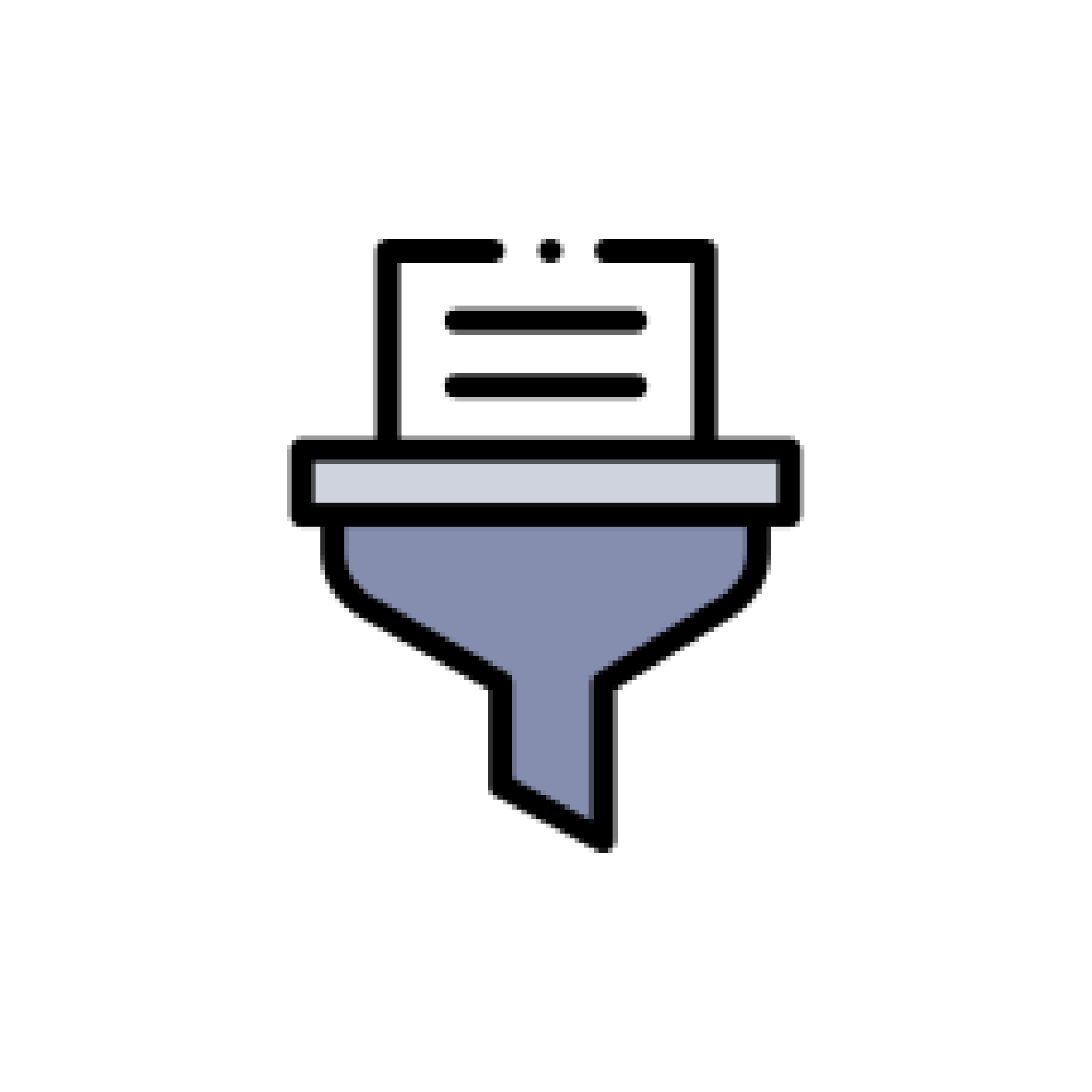}};
		\draw[thick,-stealth] (1.3, 1.1) -- (3, 1.1);
		\draw[thick,-stealth] (4.5, 1.1) -- (5.7, 1.1);

		\draw[thick,-stealth] (0, 0.6) -- (0, 0.2);
		\draw[thick,-stealth] (7.5, 0.6) -- (7.5, 0.2);

		\draw[dashed] (2.5, 0.1) -- (2.5, -3.2);

		\node at (7.6, 1.8) {\scriptsize Reduced training data (modified commits)};

		\begin{scope}[shift={(6.5,1.5)}]
			\draw[thick] (0,-0.1) -- (0.5,-0.1);
			\draw[thick] (1,-0.1) -- (1.5,-0.1);
			\draw[very thick,color=green] (0.95,-0.15) -- (1.7,-0.15);

			\draw[very thick,color=red] (-.05,-0.35) -- (0.8,-0.35);
			\draw[very thick,color=green] (0.95,-0.35) -- (1.8,-0.35);
			\draw[thick] (0,-0.4) -- (0.6,-0.4);
			\draw[thick] (1,-0.4) -- (1.6,-0.4);
		\end{scope}

		\draw[thick] (-2, .1) rectangle (2, -.9) node[pos=.5] {Large Language Model};
		\draw[thick] (5.5, .1) rectangle (9.5, -.9) node[pos=.5] {Large Language Model};

		\node[scale=0.05] (BuggyFile1) at (-1, -2)  {\includegraphics{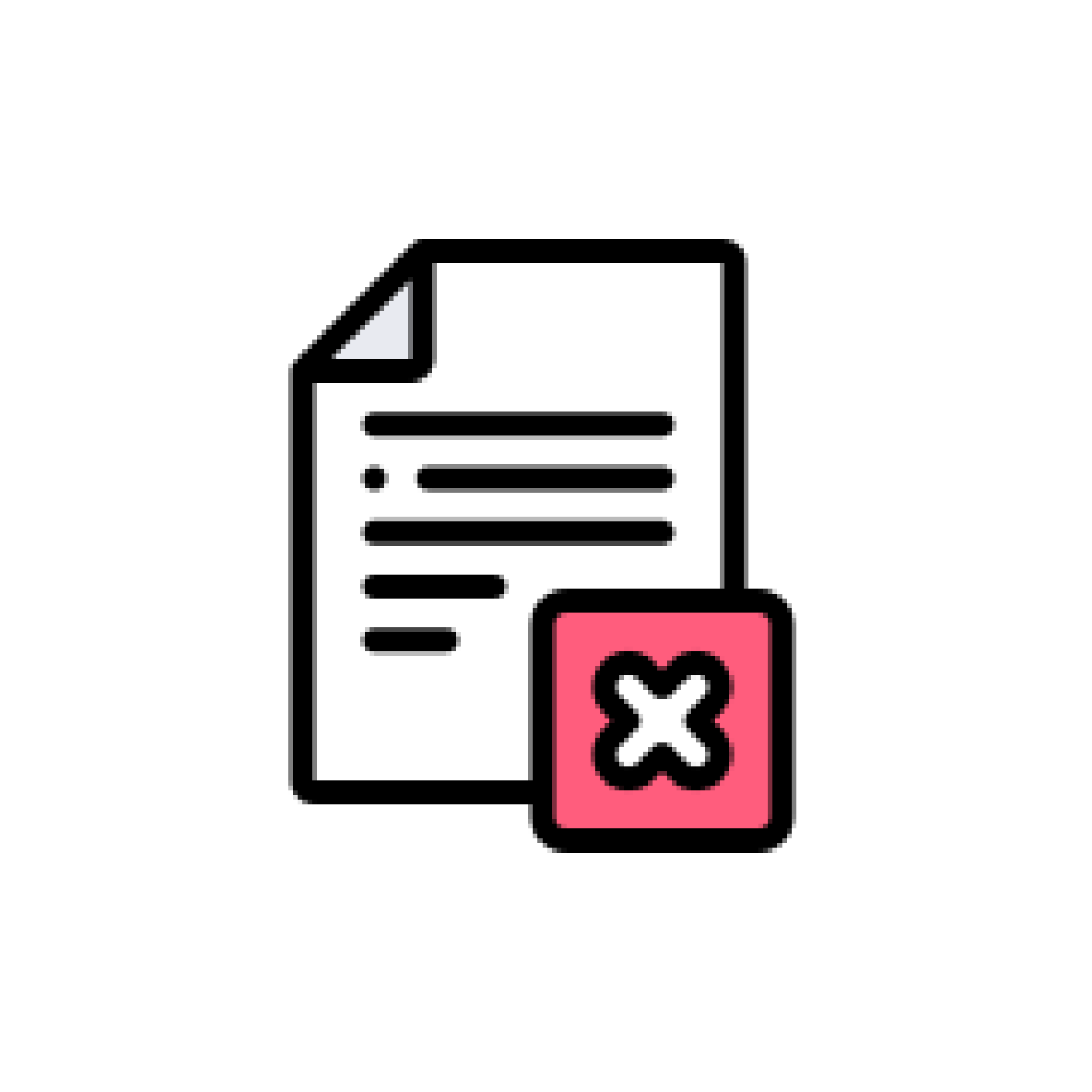}};
		\node[scale=0.05] (FixedFile1) at (1, -2)  {\includegraphics{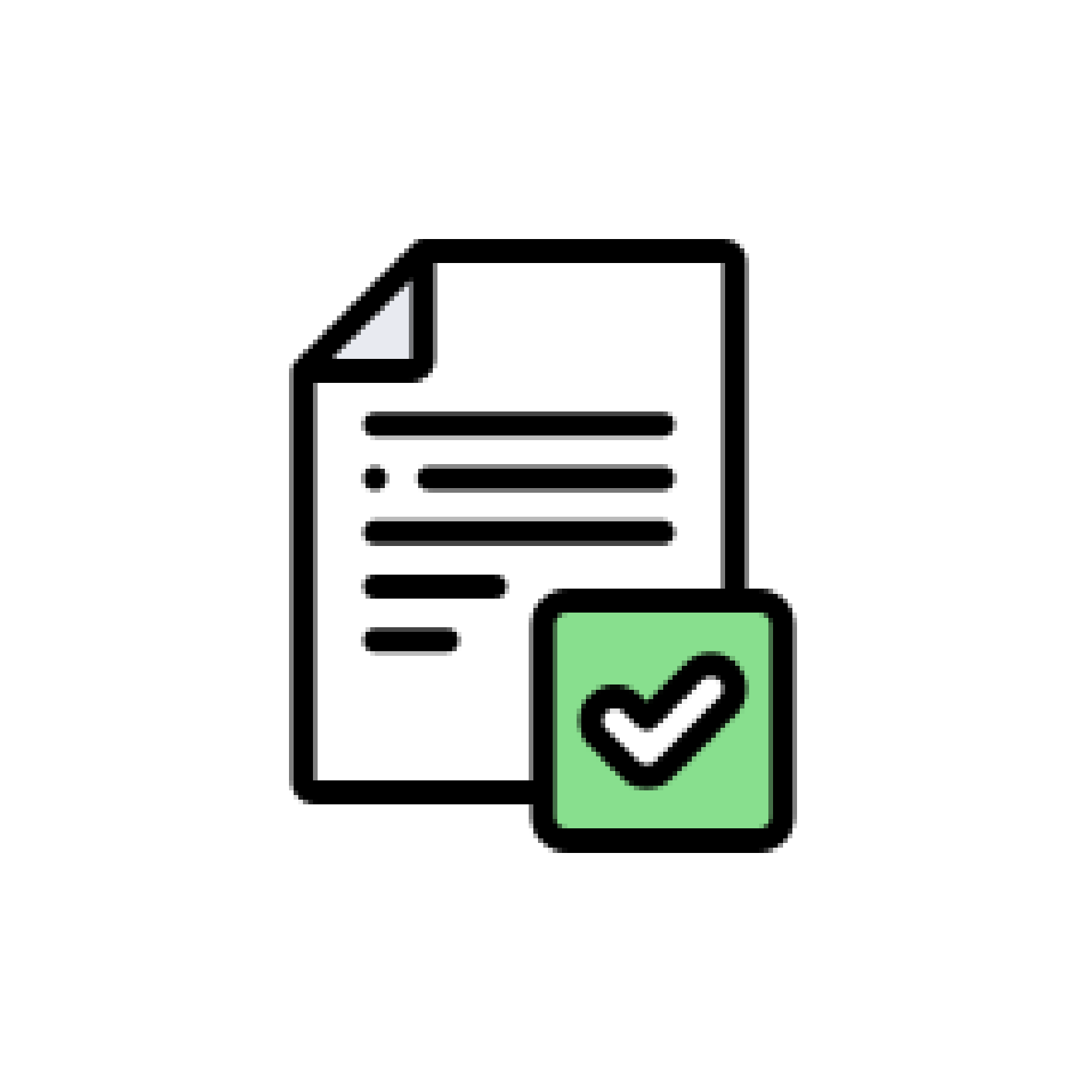}};
		\node[] (BuggyFileText) [below=-1mm of BuggyFile1] {\tiny \shortstack[c]{Code with\\Static Analysis Report}};
		\node[] (BuggyFileText) [below=-1mm of FixedFile1] {\tiny \shortstack[c]{Fixed\\Code}};

		\node[] [below=-1.5mm of CodeReduceTrain] {\tiny \shortstack[c]{\codereduce{}}};

		\node[scale=0.05] (BuggyFile2) at (3.75, -2)  {\includegraphics{figures/file_bug.pdf}};
		\node[scale=0.05] (CodeReduce2) at (5.5, -2) {\includegraphics{figures/code_reduce_icon.pdf}};
		\node[scale=0.03] (ReducedCode2) at (7, -1.7) {\includegraphics{figures/file_bug.pdf}};
		\node[scale=0.03] (Prediction2) at (8, -1.7) {\includegraphics{figures/file_fix.pdf}};
		\node[scale=0.05] (MergeBack2) at (9.5, -2) {\includegraphics{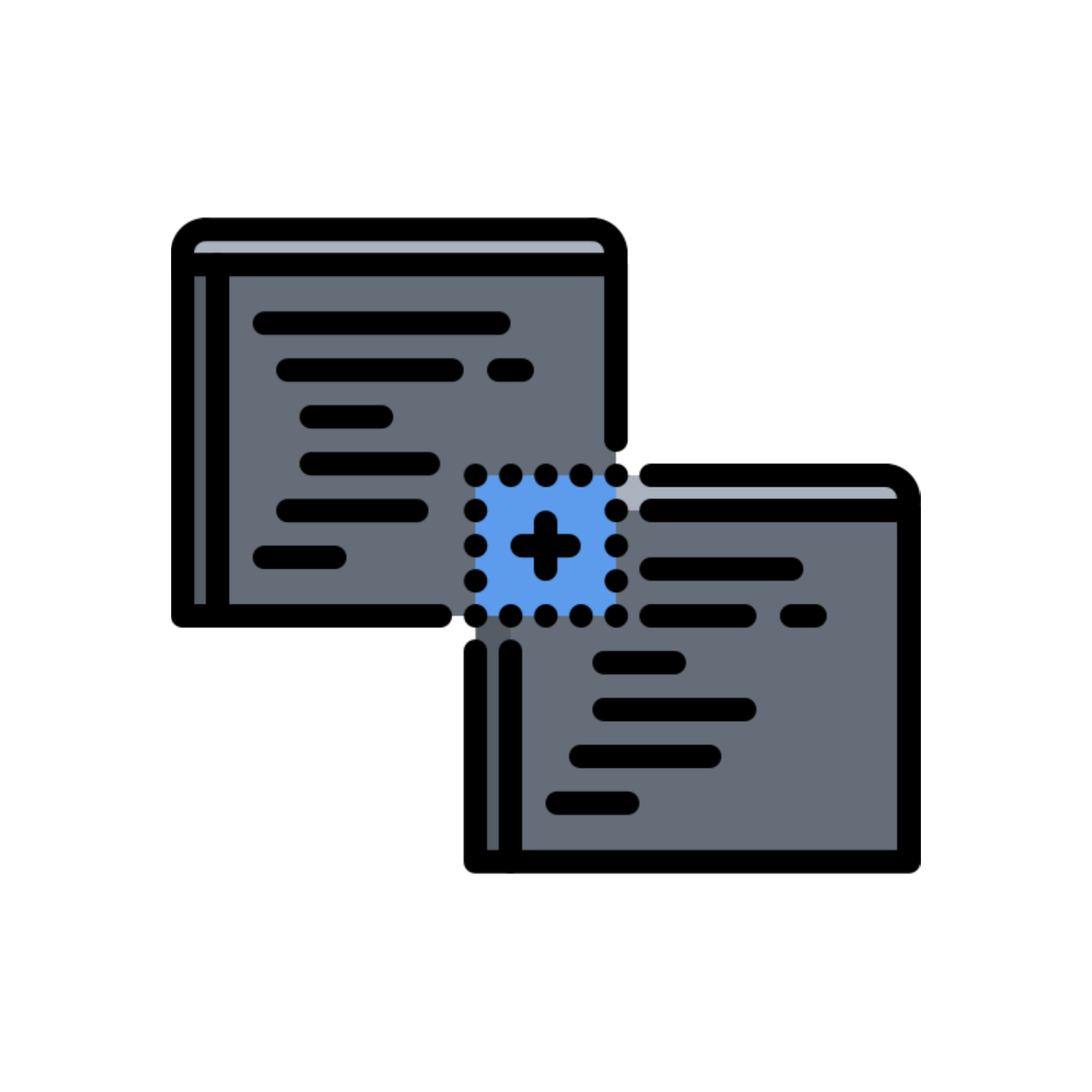}};
		\node[scale=0.05] (FixedFile2) at (11.25, -2)  {\includegraphics{figures/file_fix.pdf}};
		\node[] [below=-1mm of BuggyFile2] {\tiny \shortstack[c]{Code with\\Static Analysis Report}};
		\node[] [below=-1mm of FixedFile2] {\tiny \shortstack[c]{Fixed\\Code}};
		\node[] [below=-1mm of CodeReduce2] {\tiny \shortstack[c]{\codereduce{}}};
		\node[] [below=-1mm of MergeBack2] {\tiny \mergeback{}};

		\draw[thick, -stealth] (BuggyFile2) -- (CodeReduce2);
		\draw[thick, -stealth] (MergeBack2) -- (FixedFile2);

		\draw[thick, -stealth] (6.1, -1.75) -- (6.5, -1.75);
		\draw[thick, -stealth] (8.4, -1.75) -- (8.8, -1.75);
		\draw[thick, -stealth] (6.1, -2.25) -- (8.8, -2.25);

		\node at (0, -3.5) {(a)};
		\node at (7.5, -3.5) {(b)};

		\draw[thick,-stealth] (-0.5,-1.5) -- (-0.05,-1);
		\draw[thick,-stealth] (0.05,-1) -- (0.5,-1.5);

		\draw[thick,-stealth] (7.2,-1.3) -- (7.45,-1);
		\draw[thick,-stealth] (7.55,-1) -- (7.8,-1.3);

		\node at (.9, 0.4) {\scriptsize train/fine-tune};
		\node at (8.4, 0.4) {\scriptsize train/fine-tune};

		\node at (.9, -1.2) {\scriptsize predict};
		\node at (8.3, -1.15) {\scriptsize predict};
	\end{tikzpicture}
	\caption{Pipeline for automatic bug fixing with (a) only a large language model or (b) with complete process proposed by \tool{} combining \codereduce{} with a large language model.}\label{Fi:system}
\end{figure*}

\paragraph{Key challenge: the need to learn long-distance relationships}
At the same time, the nature of fixing semantic program errors is such that correct fixes must understand long-distance relationships in the to-be-fixed code, yet learning such complex attention mechanisms expectedly requires large amounts of training data, which, as discussed above, is practically infeasible to obtain. Even if one somehow obtains such data, the inherent limitation in the context and output size of LLMs impedes their direct utilization both in fine-tuning and few-shot learning settings. While recent models have been developed to accommodate extensive context sizes, it is important to underline that supporting such dimensions in terms of computational resources and latency does not equate to effectively learning from long-range dependencies. If not constrainted by input size, these models still encounter challenges in handling tasks with prolonged contextual requirements and also often exhibit constraints in the number of output tokens beyond a certain threshold. Overall, this means that a direct application of LLMs(even state-of-the-art) on this task, as illustrated in Figure~\ref{Fi:system} (a), is unlikely to work well.

We remark that our experiments with directly applying extensively pretrained models such as GPT-4 led to worse result on complex bugs in comparison to applying the same technique on simple few-line bugs.

\paragraph{This work: Code-Reduced Dataset \& LLM-based Code Fixing}
In this work, we address the above challenges in two steps. First, we perform extensive labeling of open-source commits and construct a clean dataset of issues and fixes for JavaScript programs. Second, we propose a new approach that improves the learning and prediction capability of code models by leveraging static code analysis, in line with prior work, which showed that program analysis features can play a significant role in the quality of the learned model, having the same quantitative effect as adding (an order of magnitude) more labeled data~\cite{DBLP:conf/pldi/RaychevVY14, programasgraph, DBLP:conf/iclr/HellendoornSSMB20}. Our technique is inspired by cReduce~\cite{DBLP:cReduce} -- a tool originally used to minimize code snippets such that a compiler bug present in the original code is preserved in the minimized snippet. Instead of compiler bugs, here we reduce the input code such that a static analysis report in the original program is still reported by the same static analyzer in the reduced code snippet. The reduced snippet contains the core of the problem, meaning that it focuses the training of the attention mechanism to very few lines of code, in turn, \textit{reducing the samples length of the required (and difficult to obtain) training data}. The code reduction technique serves as a way of context extraction that allows modeling long-range dependencies and makes the repair system independent of the original file size. Using code reduction, we obtain a new cleaned dataset, where each sample only contains code related to its particular issue. We visualize the entire process in Figure~\ref{Fi:system} (b). \codereduce{} still allows us to use models pretrained on text and code while replacing the need to learn attention over entire files or large contexts. As the samples are much shorter, a language model can effectively learn to fix bugs in programs either with fine-tuning or few-shot learning. Once a new program needs a fix for a reported issue, its code is reduced, and the LLM then fixes this reduced code. The LLMs' output is then merged back into the original file by using a patch procedure similar to how git performs code patching. Interestingly, as code reduction dramatically decreases the input length fed into the model, it even enables the usage of models of standard attention length ~\cite{t5, codet5} that can now attend to the entire reduced code snippet with a vanilla attention mechanism.

Importantly, an automatic bug-fixing tool should be immediately useful to developers inside their IDE, if the predictions can be done in real-time. \codereduce{} makes real-time serving possible as the number of tokens to be processed and to be generated by the model are significantly smaller than standard approaches. Moreover, to further improve real-time latency, we propose a speed-up over hierarchical delta debugging~\cite{DBLP:HDD} based on provenance information from the static analysis tool. This extension reduces the calls to the underlying static analyzer by $30\%$, leading to a reduced latency.

\subsection{Main Contributions}\label{sec:contrib}
\begin{itemize}[left=1mm]
	\item A static analysis-enabled novel code reduction technique which produces a small representation of the program, one that easily fits into the attention window of a model, yet contains the necessary information for learning a correct fix.
	\item An extensive evaluation of several models both with fine-tuning (\textsc{Mixtral}, \textsc{StarCoderBase}, \textsc{T5}) and few-shot learning (\textsc{GPT-3.5}, \textsc{GPT-4}) on different datasets including reduced code, small window, long context and entire file.
	\item Algorithms for reducing code while preserving a static analysis alarm and then for merging the fixed reduced code prediction back into the original source file.
	\item An end-to-end code fixing tool \tool{}\footnote{The work is available at \url{https://github.com/BBerabi/deepcode\_ai\_fix}}: substantially improving both readily available models like GPT-4 using few-shot learning, as well as fine-tuning models based on our extensive evaluations.
\end{itemize}


\section{Overview of \tool{}}\label{sec:approach-general}

In this section, we present an overview of \tool{} on a motivating example that contains a security vulnerability. Figure \ref{fig:original-pre-post} (a) shows a piece of JavaScript code that includes a server with a Path Traversal vulnerability~\cite{OWASP-path-traversal}. There are static analysis tools that can discover this vulnerability and report it on the \RoundedFrameWord{frame-highlighted} line in the code. To fit the figure, we significantly simplified the actual code example. In Figure \ref{fig:original-pre-post} (b), we show a human-provided fix for this security vulnerability in a diff format. The human fix does not only contain the removal of the vulnerability but also includes an unrelated change \texttt{\TightColoredWord{myred}{doBBB()}$\;\to\;$\TightColoredWord{mygreen}{doXYZ()}}.


\begin{figure*}
	\centering
	\begin{subfigure}{.998\columnwidth}
		\begin{lstlisting}[language=js, title={(a) Input: vulnerable pre-version}, captionpos=t, numberstyle=\linecolors{white}{0}{0}]
const fs = require('fs');
const path = require('path');
function doUnrelated() {
  doAAA();
(*\HL{myred}*)- doBBB();
  doCCC();
}
function uploadFile(fileName) {
  doDDD();
(*\HL{myemptygray}*)
  dest = path.join("www", fileName);
(*\RoundedFrameLine*)  fs.createWriteStream(dest);
}
function serverHandler(request, reply) {
  doUnrelated();
  uploadFile(request.payload.fileName);
}
    \end{lstlisting}
	\end{subfigure}\hfill
	\begin{subfigure}{.998\columnwidth}
		\begin{lstlisting}[language=js,  title={(b) Input: non-vulnerable post-version}, captionpos=t, numberstyle=\linecolors{white}{0}{0}]
const fs = require('fs');
const path = require('path');
function doUnrelated() {
  doAAA();
(*\HL{mygreen}*)+ doXYZ();
  doCCC();
}
function uploadFile(fileName) {
  doDDD();
(*\HL{mygreen}*)+ fileName = path.basename(fileName);
  dest = path.join("www", fileName);
  fs.createWriteStream(dest);
}
function serverHandler(request, reply) {
  doUnrelated();
  uploadFile(request.payload.fileName);
}
	  \end{lstlisting}
	\end{subfigure}

	\vspace{-1mm}
	\begin{subfigure}{.998\columnwidth}
		$$\downarrow\text{\codereduce}\downarrow$$
	\end{subfigure}\hfill
	\begin{subfigure}{.998\columnwidth}
		$$\downarrow \downarrow$$
	\end{subfigure}
	\vspace{2mm}

	\begin{subfigure}{.998\columnwidth}
		\begin{lstlisting}[language=js, title={(c) Reduced pre-version}, numberstyle=\linecolors{white}{0}{0}]
const fs = require('fs');
const path = require('path');
function uploadFile(fileName) {
(*\HL{myemptygray}*)
  dest = path.join("www", fileName);
(*\RoundedFrameLine*)  fs.createWriteStream(dest);
}
function serverHandler(request, reply) {
  uploadFile(request.payload.fileName);
}
    \end{lstlisting}
	\end{subfigure}\hfill
	\begin{subfigure}{.998\columnwidth}
		\begin{lstlisting}[language=js, title={(d) Corresponding reduced post-version}, numberstyle=\linecolors{white}{0}{0}]
const fs = require('fs');
const path = require('path');
function uploadFile(fileName) {
(*\HL{mygreen}*)+ fileName = path.basename(fileName);
  dest = path.join("www", fileName);
  fs.createWriteStream(dest);
}
function serverHandler(request, reply) {
  uploadFile(request.payload.fileName);
}
	  \end{lstlisting}
	\end{subfigure}

	\vspace{-1mm}
	\begin{subfigure}{.8\columnwidth}
		$$\downarrow\text{\small\mergeback}\downarrow$$
	\end{subfigure}
	\vspace{2mm}

	\begin{subfigure}{.998\columnwidth}
		\begin{lstlisting}[language=js, title={The original vulnerable code (repeated for clarity)}, numberstyle=\linecolors{white}{0}{0}]
const fs = require('fs');
const path = require('path');
function doUnrelated() {
  doAAA();
  doBBB();
  doCCC();
}
function uploadFile(fileName) {
  doDDD();
(*\HL{myemptygray}*)
  dest = path.join("www", fileName);
(*\RoundedFrameLine*)  fs.createWriteStream(dest);
}
function serverHandler(request, reply) {
  doUnrelated();
  uploadFile(request.payload.fileName);
}  
    \end{lstlisting}
	\end{subfigure}\hfill
	\begin{subfigure}{.998\columnwidth}
		\begin{lstlisting}[language=js, title={(e) Automatically fixed version of the original code}, numberstyle=\linecolors{white}{0}{0}]
const fs = require('fs');
const path = require('path');
function doUnrelated() {
  doAAA();
  doBBB();
  doCCC();
}
function uploadFile(fileName) {
  doDDD();
(*\HL{mygreen}*)+ fileName = path.basename(fileName);
  dest = path.join("www", fileName);
  fs.createWriteStream(dest);
}
function serverHandler(request, reply) {
  doUnrelated();
  uploadFile(request.payload.fileName);
}
	  \end{lstlisting}
	\end{subfigure}
	\vspace{-2mm}
	\caption{An example run of our data creation pipeline on a diff containing a path traversal vulnerability~\cite{OWASP-path-traversal} fix (a)$\to$(b). \textbf{Colors:} \TightColoredWord{myred}{red -}/\TightColoredWord{mygreen}{green +}/\TightColoredWord{myemptygray}{gray} represent standard diff notation, and\; \RoundedFrameWord{framed line} marks the location of the vulnerability. \codereduce{} is applied on (a), and a similar transformation is then applied on (b) to get (d), whereas \mergeback{} is applied to obtain (e) from (a) and the diff of (c)$\to$(d). Other real-world fix examples generated fixes by \tool{} can be found in Appendix \ref{sec:appendix_examples}.
	}
	\label{fig:original-pre-post}
\end{figure*}

Since the static analysis tool discovers the path traversal vulnerability in the code in (a) and does not discover it in the code in (b), this is a signal that the pair (a)$\to$(b) qualifies as a potential training data sample for learning. To obtain an even cleaner data sample, \tool{} calls the \codereduce{} component, which on its own calls the static analyzer multiple times until it produces the code shown in Figure \ref{fig:original-pre-post}~(c). Note that while the code is reduced, we discover which of the lines are essential for detecting the vulnerability: here, one needs to include the file system library to keep the server handler, and one needs to keep the data flow from the request to the file system operation. Note that there are multiple trivial but wrong ways to fix the code snippet in~Figure~\ref{fig:original-pre-post}~(c). For example, if any of its statements is deleted from this input, the vulnerability will disappear, however the program will also break its semantics in a significant way. This observation is important because it highlights that relying on simple heuristics to obtain training data is potentially dangerous. It can lead to models that trivially satisfy bug removal metrics but actually learn incorrect fixes.

To gain a useful fix sample, we take any change between (a) and (b) such that the line of this change is also included in (c) or adjacent to a line included in (c). Then we apply these changes to (c) and construct a sample, as shown in Figure \ref{fig:original-pre-post} (d). Next, we validate that the provided fix parses and removes the static analysis alarm. Let this pair of code snippets (c) and (d) be denoted ${\color{red}\texttt{REDUCED}_\texttt{PRE}}$ and ${\color{teal}\texttt{REDUCED}_\texttt{POST}}$, respectively. Let the static analysis report type be denoted \texttt{\color{purple}RULE}/\texttt{\color{blue}DESCRIPTION} -- e.g. in the figure, the rule is \texttt{PT} identifying path traversal and description contains a short text about the vulnerability.
Then, a large language model can be provided code fixing examples similar to (c) and (d) using the prompt:
\[
	\texttt{fix }\texttt{\color{purple}RULE }\texttt{\color{blue}DESCRIPTION}\texttt{ : }{\color{red}\texttt{REDUCED}_\texttt{PRE}}
\]
be fine-tuned to generate ${\color{teal}\texttt{REDUCED}_\texttt{POST}}$ given such a prompt. Note that because (c) and (d) are short code samples, they are expected to fit in the attention window of a text transformer and the learning can be efficient even with relatively few training samples. The same prompt structure can be used for inference as well.

Assume that the LLM has learned to perform fixes on programs like (c). If we provide the input (a) as a query, \tool{} will reduce the sample to program (c), the language model will transform it into (d), and then we need to merge back the resulting reduced fix program into the input program (a). This step, called \mergeback{}, is shown in~Figure~\ref{fig:original-pre-post} (e), and works by applying the changes observed between (c) and (d) onto the program (a).

It is a known problem that bug-fixing commits often contain patches irrelevant to the applied fix, such as refactorings \cite{BugBuilder}. Note that the procedure in Figure \ref{fig:original-pre-post} is also useful for data cleaning. The first pair (a)$\to$(b) is transformed into a pair (a)$\to$(e) such that the unrelated change in the first pair like \texttt{\TightColoredWord{myred}{doBBB()}$\;\to\;$\TightColoredWord{mygreen}{doXYZ()}} is removed. This side benefit of \codereduce{} plays an important role for obtaining a clean dataset.

\section{Core Components of \tool{}}\label{sec:approach-detailed}
Next, we describe \tool{}'s components.

\subsection{Code Analysis}\label{sec:code-analysis}
For code analysis, \tool{} uses Snyk Code~\cite{SnykCode} -- a proprietary commercial static analyzer that operates on source code. In this work, we only evaluate with JavaScript/TypeScript analysis, along with its extensions such as ReactJS and VueJS. As of January 2024, the analyzer implements \NumRules{} different checks that are in Appendix~\ref{sec:rule-scope}. We classified them into five categories as follows:

\AST{}:
Most of the \AST{} checks are ones that can be performed only on the abstract syntax tree (AST). Many of these rules enforce properties about the VueJS or React extensions of the language and include checks such as missing tags, duplicate variable names and patterns that usually do not depend on analyzing the control or dataflow of the program.

\Local:
The \Local{} checks use interprocedural analysis to discover that certain incorrect values would flow into methods that would not accept them, to discover that resources are not deallocated or that certain expressions have no effect or always produce results with unexpected semantics.

\FileWide:
The \FileWide{} checks also depend on interprocedural analysis, but in contrast to the \Local{} checks, the properties that they check always stay in different functions or methods in the program. These rules include checks for mismatches between the signature of a method, its implementation or its usage.

\SecurityLocal:
Similar to the \Local{} checks, \SecurityLocal{} checks verify incorrect API usage from the security perspective as well as other types of security misconfigurations, typically by tracking the set of all method calls on an object or by tracking the values passed as parameters to method or function calls.

\SecurityFlow:
The \SecurityFlow{} checks are usually the most complex checks supported by the selected static analyzer. This rule category includes all taint analysis rules that involve interprocedural dataflow analysis starting from a data source as well as the detection of various data sanitization patterns.

\textbf{Approximate Provenance Information.}
In addition to reports, static analyzers typically provide proofs of the reports, such as data flow or a flow describing the discovered defect. These, however, target human readability and are usually not the full set of program statements a static analyzer needs to reproduce the alarm. As a result, many existing analyzers provide approximate provenance that is either over-, or under- approximation of the precise provenance. Yet, even an approximation is useful to speed-up the \codereduce{} algorithm. We assume the existence of a procedure that returns the approximate set of abstract syntax tree (AST) nodes required for a report. We define a function $\textsc{ApproxProvenanceNodes}$ that takes a list of tree nodes, a static analysis report and returns a list of nodes that participate in the derivation of the static analysis report.

\subsection{\codereduce}\label{sec:codereduce}
Given a file and a vulnerability detected by our static analyzer, \tool{} calls \codereduce{} to reduce the source code to a small snippet that still reports that static analysis alarm. \codereduce{} is inspired by cReduce and delta debugging, which execute a set of transformation passes on the tree representation of the input code while checking that a detected property remains present in all of its transformation phases. In our case, all our transformations are performed on an AST and we base our algorithm on Hierarchical Delta Debugging (HDD)~\cite{DBLP:HDD}.

\begin{algorithm}
	\caption[\codereduce]{\codereduce{} procedure based on HDD~\cite{DBLP:HDD}}
	\label{algo:codereduce}
	\begin{algorithmic}[1]
		\footnotesize
		\INPUT{ $ $ \\
			$C$: source code; \\
			$\mathit{tree}$: \AST{} of the input code including the node locations in $C$; \\
			$\mathit{report}$: the vulnerability and the \AST{} node on which it was reported; \\
		}

		\OUTPUT{ $ $ \\
			$c$: reduced code; \\
			$\linemapping{}$: line mapping; \\
		}

		\STATE $\mathit{level} \gets 0$
		\STATE $\mathit{nodes{} \gets \textsc{GetNodes}(\mathit{tree}, \mathit{level})}$\label{line:reduciblenodes}
		\STATE $c \gets C$
		\WHILE {$\mathit{nodes} \neq \emptyset$} \label{line:whileloop}

		\STATE $\mathit{nodes}^\text{prov} \gets \textsc{ApproxProvenanceNodes}(\mathit{nodes}, \mathit{report})$ \label{line:provenance}
		\STATE $c^\text{prov}, \mathit{tree}^\text{prov}, \linemapping{}^\text{prov} \gets
			\textsc{RemoveNodesFromCode}(\mathit{nodes} \setminus  \mathit{nodes}^\text{prov}, c, \mathit{tree}, \linemapping{})$

		\IF {$\mathrm{Parses}(c^\text{prov})$ $\wedge$ $\mathrm{ReportExists}(c^\text{prov}, \mathit{report})$} \label{line:checkreport}
		\STATE $\mathit{nodes} \gets \mathit{nodes}^\text{prov}$ \label{line:continuenodes}
		\STATE $c, \mathit{tree}, \linemapping{} \gets c^\text{prov}, \mathit{tree}^\text{prov}, \linemapping{}^\text{prov}$ \label{line:updatecode}
		\ENDIF
		\STATE $c, \mathit{tree}, \linemapping{} \gets \ddmin{}(\mathit{nodes}, \mathit{report}, c, \mathit{tree}, \linemapping{})$ \label{line:dd}
		\STATE $\mathit{level} \gets \mathit{level} + 1$ \label{line:levelup}
		\STATE $\mathit{nodes} \gets \textsc{GetNodes}(\mathit{tree}, \mathit{level})$ \label{line:nextnodes}
		\ENDWHILE
	\end{algorithmic}
\end{algorithm}

\begin{figure*}
	\vspace{-2mm}
	\centering
	\begin{subfigure}{.95\columnwidth}
		\begin{lstlisting}[language=jssmall, title={Original Code}]
const express = require('express');
const fs = require('fs');
express().post('/path', (req, res) => {
    var options = {
        dotfiles: fs.ReadFileSync('cfg')
    };
    res.sendFile(req.params.name, options);
});
    \end{lstlisting}
	\end{subfigure}\hfill
	\begin{subfigure}{.04\columnwidth}
		\begin{lstlisting}[language=jssmall, title={$l_{C \to c}$}]
(*\LessTightColoredWord{myyellow}{1}*) (*\LessTightColoredWord{myyellow}{1}*)
(*\LessTightColoredWord{myblue}{2}*) (*\phantom{\LessTightColoredWord{mypurple}{X}}*)
(*\LessTightColoredWord{mybrown}{3}*) (*\LessTightColoredWord{mybrown}{2}*)
(*\LessTightColoredWord{myorange}{4}*) (*\phantom{\LessTightColoredWord{mypurple}{X}}*)
(*\LessTightColoredWord{myorange}{5}*) (*\phantom{\LessTightColoredWord{mypurple}{X}}*)
(*\LessTightColoredWord{myorange}{6}*) (*\phantom{\LessTightColoredWord{mypurple}{X}}*)
(*\LessTightColoredWord{mypurple}{7}*) (*\LessTightColoredWord{mypurple}{3}*)
(*\LessTightColoredWord{mypurple}{8}*) (*\LessTightColoredWord{mypurple}{4}*)
    \end{lstlisting}
	\end{subfigure}\hfill
	\begin{subfigure}{.95\columnwidth}
		\begin{lstlisting}[language=jssmall, title={Reduced Code}]
const express = require('express');
(*\HL{myemptygray}*)
express().post('/path', (req, res) => {
(*\HL{myemptygray}*)
(*\HL{myemptygray}*)
(*\HL{myemptygray}*)
    res.sendFile(req.params.name, options);
});
    \end{lstlisting}
	\end{subfigure}
	\vspace{-4mm}
	\caption{An example code reduction. The usage of the variable \texttt{options} remains in the reduced code, but its definition is deleted. The definition of \texttt{express} is not deleted because it is an import statement related to the vulnerability. The reduced code is not executable, but the analyzer can detect the path traversal vulnerability in both versions.}
	\label{fig:codereduce}
\end{figure*}

HDD is itself based on a Delta Debugging~\cite{DBLP:DD} procedure \ddmin{} that is called for the nodes at each level of the AST of the currently reduced code. \ddmin{} takes as input a set of program elements (in our case AST nodes), a static analysis report and the code with its tree. Its output is a further reduced code with a reduced tree that still keep the static analysis alarm present in the program. \ddmin{} performs multiple attempts to remove tree nodes and calls the underlying static analysis multiple times to check the presence of the report. We refer to \cite{DBLP:DD} for more details.

We modify the HDD procedure to take into account the approximate provenance information the static analysis alarm provides. Algorithm \ref{algo:codereduce} describes the concrete procedure. It begins at the root node of the \AST{}. At each level, the algorithm filters out the nodes that should be considered for deletion depending on the level in the tree (line \ref{line:reduciblenodes}). The provenance handling is shown in lines \ref{line:provenance}-\ref{line:updatecode}. The algorithm attempts to remove all nodes from the level that were not part of the approximate provenance and if the report is not removed (line \ref{line:checkreport}), the removal is accepted. The remaining lines \ref{line:dd}-\ref{line:nextnodes} continue as in the original HDD algorithm with a call to \ddmin{} on the nodes in the current level and moving to the next level. To maintain guarantees also when the provenance is approximate, the algorithm still performs the \ddmin{} procedure regardless of the success of the provenance nodes removal step.

To make \codereduce{} useful for the automatic bugfixing case, we also modify $\textsc{GetNodes}$ to only delete statements that are complete lines and avoid deleting lines only partially because partially deleted lines are harder to merge back into the original code. Additionally, we only consider nodes that delete statements and declarations but keep expressions inside statements. This means that for example, statements such as \texttt{call(arg1, arg2)} will not be reduced to \texttt{call()}, \texttt{call(arg1)} or \texttt{call(arg2)}.

When used as part of a loop until fixpoint, Algorithm \ref{algo:codereduce} has the same 1-tree-minimality guarantees as in \cite{DBLP:HDD}. This means the returned code cannot be further reduced by removing any single node in the tree, such that the static analysis report remains present. The reason for a fixpoint loop is that a node on a higher level might become deletable only after having another node deleted on a lower level.

\subsection{\mergeback}

Even though \codereduce{} brings many advantages, it comes with the cost of having predictions that are not directly usable. The reason is that the predictions, just like the model input, are also in a reduced form. Hence, to truly provide end-to-end bug fixing, \tool{} must merge the generated code back into the original file.

We present the \mergeback{} procedure in Algorithm \ref{algo:mergeback}. It starts by comparing the reduced input code $c$ with the predicted fix $p$. This comparison uses \gitdiff{} and aims to compute a one-to-many mapping \replacementmapping{} between the lines of $c$ and $p$ (line \ref{line:replacement}). For each line $l_c$ in the reduced code $c$, \replacementmapping{} contains a potentially empty set of lines from the prediction $p$, by which the line $l_c$ should be replaced. Technically, if $l_c$ is mapped to $l_p$, we write that as $(l_c, l_p) \in \replacementmapping{}$ and we assume iterations over this set return the elements in sorted order. Once the mappings are computed, the original non-reduced file $\mathcal{C}$ and the fix $p$ are split into lines, and $\fixedlines$ is initialized to an empty sequence of strings (line \ref{line:init}).

\begin{algorithm}
	\caption[\mergeback]{Merging prediction back into the original file}
	\label{algo:mergeback}
	\begin{algorithmic}[1]
		\footnotesize
		\INPUT{$ $ \\
			$C$: source code; \\
			$c$: reduced code; \\
			$p$: prediction on reduced code;\\
			$\linemapping{}$: line mapping obtained from running \codereduce{}; \\
		}

		\OUTPUT{$ $ \\
			$C''$: resulting source code; \\
		}

		\STATE $\replacementmapping{} \gets \text{ComputeReplacementMapping}(c, p)$ \label{line:replacement}
		\STATE \sourcelines{} $\gets \text{SplitLines}(C)$ \; \plines{} $\gets \text{SplitLines}(p)$ \; \\
		\fixedlines{} $\gets \langle \; \rangle $ \label{line:init} \;

		\FOR {$l_C \in \sourcelines{}$} \label{line:iteratelines}
		\IF {$\exists\, l_c \colon (l_c, l_C) \in \linemapping$}  \label{line:checkinreduced}
		\STATE {$\fixedlines{} \gets \fixedlines \cup \langle l_p \in \plines{} \;\text{s.t.}\; (l_c, l_p) \in \replacementmapping \rangle$} \label{line:addfromreplacement}
		\ELSE
		\STATE {$\fixedlines{} \gets \fixedlines \cup \langle l_C \rangle$}   \label{line:copyfromsource}
		\ENDIF
		\ENDFOR

		\STATE  {\bfseries return} \fixedlines{}$.\mathrm{join}(\text{``\textbackslash n''})$ \label{line:joinandreturn}
	\end{algorithmic}
\end{algorithm}

\begin{figure*}
	\vspace{-2mm}
	\centering
	\begin{subfigure}{.95\columnwidth}
		\begin{lstlisting}[language=jssmall, title={Reduced code $c$}]
import express from 'express';
(*\HL{myemptygray}*)
import bodyParser from 'body-parser';
const app = express();
(*\HL{myred}*)app.disable('x-powered-by');

app.use(bodyParser.urlencoded());
    \end{lstlisting}
	\end{subfigure}\hfill
	\begin{subfigure}{.04\columnwidth}
		\begin{lstlisting}[language=jssmall, title={\replacementmapping{}}]
(*\LessTightColoredWord{myyellow}{1}*) (*\LessTightColoredWord{myyellow}{1}*)
(*\LessTightColoredWord{myemptygray}{1}*) (*\LessTightColoredWord{mygreen}{2}*)
(*\LessTightColoredWord{myyellow}{2}*) (*\LessTightColoredWord{myyellow}{3}*)
(*\LessTightColoredWord{myyellow}{3}*) (*\LessTightColoredWord{myyellow}{4}*)
(*\LessTightColoredWord{myred}{4}*) (*\LessTightColoredWord{mygreen}{5}*)
(*\LessTightColoredWord{myred}{4}*) (*\LessTightColoredWord{mygreen}{6}*)
(*\LessTightColoredWord{myyellow}{5}*) (*\LessTightColoredWord{myyellow}{7}*)
    \end{lstlisting}
	\end{subfigure}\hfill
	\begin{subfigure}{.95\columnwidth}
		\begin{lstlisting}[language=jssmall, title={Prediction $p$}]
import express from 'express';
(*\HL{mygreen}*)import helmet from 'helmet';
import bodyParser from 'body-parser';
const app = express();
(*\HL{mygreen}*)app.use(helmet());
(*\HL{mygreen}*)app.use(helmet.hsts());
app.use(bodyParser.urlencoded());
    \end{lstlisting}
	\end{subfigure}
	\vspace{-4mm}
	\caption{An example illustrating diff hunks between reduced code $c$ and prediction $p$ and the computed mapping $\replacementmapping{}$ between them.}
	\label{fig:mergeback}
\end{figure*}

After that, the algorithm iterates over the lines in the non-reduced input \sourcelines{} (line \ref{line:iteratelines}) and checks for each line whether it is used in the reduced code or not (line \ref{line:checkinreduced}). If a line does not participate in the reduced code $c$, it implies that the line is not important for the bug fix and the model does not see this line while generating the prediction. Therefore, that line must be kept as is, so we append this into \fixedlines{} (line \ref{line:copyfromsource}). In the other case, the prediction must decide how the line should be replaced. Therefore, we look up in \replacementmapping{} to obtain the list of prediction lines meant to replace the current line and add each of them to \fixedlines{} (line \ref{line:addfromreplacement}). In the end, the algorithm joins the fixed lines \fixedlines{} by the new line character and returns the result (line \ref{line:joinandreturn}). Finally, the algorithm depends on computing \replacementmapping{}. This is done by essentially doing a \gitdiff{} and obtaining the so called "hunks" of corresponding mapped pieces of code. For each hunk, we calculate the corresponding line mapping as visualized in Figure \ref{fig:mergeback}.

\section{Data Collection}\label{sec:data}

In this section, we investigate the problem of collecting commit data and cleaning it up for building a repair tool. Our dataset is built from a large-scale crawl of GitHub that includes the top 500k repositories sorted by the number of stars. From these repositories, we selected all commits that do not merge branches, do not do file renaming, affect JavaScript and both versions of the code parse correctly and have appropriate code license. This resulted in a dataset of $6$ million commits. We applied the static analyzer on all commits and obtained around $380$ thousand (pre-, post-) file pairs that fixed a static analysis report -- this means that a certain report was present in the pre-version of the pair and not reported anymore at the corresponding location in its post-version. The statistics about fixes are summarized per report type in the second column of Table~\ref{table:random_commits}. Note that this number of samples is still overly optimistic about the amount of training data usable for learning.

\newcounter{localcolcounter}
\newcommand{\CC}{\stepcounter{localcolcounter}\shortstack[r]{\thelocalcolcounter.\,\\\phantom{X}}}
\begin{table*}
	\caption{Fix statistics on randomly sampled commits that remove static analysis reports.}\label{table:random_commits}
	\centering
	\footnotesize
	\begin{tabular}{r@{\,}lr@{\,}lr@{\,}lr@{\,}lr@{\,}lr@{\,}lr@{\,}l}
		\toprule
		\CC & \shortstack[l]{Issue                                                                            \\Category} & \CC & \shortstack[l]{Candidate\\Commits} & \CC & \shortstack[l]{Good\\Fix} & \CC &  \shortstack[l]{Deletion/\\Refactor} & \CC & \shortstack[l]{Accidental\\Fix} & \CC & \shortstack[l]{Precision\\Loss} & \CC & \shortstack[l]{Other\\\phantom{XX}} \\
		\midrule
		    & \AST                 &  & $106\,386$ &  & $38\%$ &  & $56\%$ &  & $6\%$  &  & $0\%$  &  & $0\%$ \\
		    & \Local               &  & $154\,945$ &  & $26\%$ &  & $56\%$ &  & $12\%$ &  & $4\%$  &  & $2\%$ \\
		    & \FileWide            &  & $35\,206$  &  & $36\%$ &  & $32\%$ &  & $24\%$ &  & $4\%$  &  & $4\%$ \\
		    & \SecurityLocal       &  & $26\,499$  &  & $18\%$ &  & $70\%$ &  & $8\%$  &  & $4\%$  &  & $0\%$ \\
		    & \SecurityFlow        &  & $58\,523$  &  & $4\%$  &  & $78\%$ &  & $8\%$  &  & $10\%$ &  & $0\%$ \\
		\bottomrule
	\end{tabular}
	\\[1mm]
\end{table*}
\let\CC\undefined

To check the data quality, we randomly sampled $50$ file pairs removing reports from each of the five categories and manually evaluated if the change contains a proper fix for the issue at hand. Our findings are in columns 3-7 of Table~\ref{table:random_commits}. Overall, the table shows that a significant amount of the file pairs have fixed the issue in a way that a model cannot and should not utilize for learning. The most frequent problem is that instead of a good fix (column 3), the code is deleted or refactored (column 4).

A more interesting observation, though, was that as the complexity of a static analysis report increases, the probability of observing a correct fix drops significantly. Security issues are often left unfixed in code but are frequently removed or moved to another location as part of a refactoring. We also observed some cases of accidental fixes (column 5), where the affected code was turned into dead code, or an unrelated change removed a particular report. Besides, the static analyzer is also not bullet-proof precision-wise. A few instances resulted in approximations that caused the analyzer to report false positives or under-approximations that caused the analysis to drop the report although no proper fix was applied (column 6). For some issues, we could observe incorrect fixes, but we could not attribute the sample to one of these problems (column 7). These results also show that as opposed to heuristic data cleaning applied in prior works such as TFix~\cite{TFix} that worked on simple issues and Lint warnings, the more complex static analysis reports would also need more manual labeling effort to take into account the complexities in the input data. Based on this finding, we designed our data collection pipeline accordingly and manually labelled the candidate file pairs in a web-application we developed.

\begin{figure*}
	\hspace{2cm}
	\begin{tikzpicture}
		\scriptsize

		\tikzstyle{arrow} = [thick,->,>=stealth]

		\node[draw, rectangle, thick] (git_commits) at (-18, 3) {\shortstack{Git Commits\\$(C, C')$}};
		\node[draw, rectangle, thick] (candidate_commits) at (-13, 3) {\shortstack[c]{\shortstack{Candidate Commits\\$(C, C')$}}};
		\node[draw, rectangle, thick] (full_original) at (-9, 3) {\shortstack[c]{\JsOriginalData{}\\ $(C, C')$}};

		\node[draw, thick, rectangle, minimum width=3.5cm] (code_reduced) at (-4.5, 3.5) {\shortstack[c]{\tiny \JsReducedData{} \tiny $(C, C') \to (c, c')$}};
		\node[draw, thick, rectangle, minimum width=3.5cm] (long_context) at (-4.5, 3.0) {\shortstack[c]{\tiny \JsWindowFiftyData{} \tiny $(C, C') \to (c'', c''')$}};
		\node[draw, thick, rectangle, minimum width=3.5cm] (window3) at (-4.5, 2.5) {\shortstack[c]{\tiny \JsWindowThreeData{} \tiny $(C, C') \to (\hat{c}'', \hat{c}''')$}};

		\def\mymargin{0.2cm}

	\draw[->] ($(git_commits.east) + (\mymargin, 0)$) -- ($(candidate_commits.west) - (\mymargin,0)$) node [midway, above] {\shortstack{Analysis: issue $I$}} node[midway, below] {on line $C{:}\ell$ fixed in $C'$};
	\draw[->] ($(candidate_commits.east) + (\mymargin, 0)$) -- ($(full_original.west) - (\mymargin,0)$) node [midway, above] {\shortstack{Manual}} node [midway, below] {Labelling};

		\draw[->] ($(full_original.east) + (\mymargin, 0.1)$) -- ($(code_reduced.west) - (\mymargin,0)$);
		\draw[->] ($(full_original.east) + (\mymargin, 0)$) -- ($(long_context.west) - (\mymargin,0)$);
		\draw[->] ($(full_original.east) + (\mymargin, -0.1)$) -- ($(window3.west) - (\mymargin,0)$);

	\end{tikzpicture}
	\caption{Overview of the data pipeline. See the main body for a detailed explanation of the steps.}
	\label{fig:data-pipeline}
\end{figure*}

\textbf{Data Collection \& Cleaning.}
We created data according to the pipeline shown in~Fig.~\ref{fig:data-pipeline}: For each pair of pre-/post-commit source files $(C, C')$ we ran the static analysis and collected the candidate file pairs where a certain issue $I$ present on line $\ell$ in $C$ was not reported anymore in $C'$. Then, we manually labelled candidate samples as negative and positive with respect to making a proper code change fixing the issue. We call the output of such a process a \textit{fix pair}. On the cleaned fix pairs, we applied \codereduce{} for the issue $(C, \ell, I)$. This step's output, apart from yielding the reduced pre-commit code $c$, allows obtaining a \textit{reduced version $c'$ of the post-commit}, thus resulting in the reduced pair $(c, c')$. We generate final data in several flavours: \JsOriginalData{}, \JsReducedData{}, \JsWindowFiftyData{}, and \JsWindowThreeData{}. Below we give explanations and the motivation behind each flavour.
\begin{table*}
	\centering
	\renewcommand{\arraystretch}{1.2}
	\caption{Data statistics, including token size percentiles. Tokenization was performed using the tokenizer of StarCoderBase~\cite{starcoder} model. Note that \codereduce{} makes data suitable even for feeding into transformers with short attention range.}
	\footnotesize
	\begin{tabular}[c]{@{\quad}l*{7}{>{\centering\arraybackslash}p{\widthof{XXXXX}}}@{\quad}}
		\toprule
		\mulro[.6]{2}{Dataset} & \multicolumn{2}{c}{\# Datapoints} & \multicolumn{5}{c}{Token Size @ $p$-Percentile}                                               \\

		\cmidrule(l{1pt}r{2pt}){2-3} \cmidrule(l{2pt}r{1pt}){4-8}
		\bottomrule
		                       & Train                             & Test                                            & $p=25$ & $p=50$ & $p=75$ & $p=95$ & $p=99$  \\
		\midrule
		\JsOriginalData        & \mulro{4}{$3532$}                 & \mulro{4}{$1818$}                               & $438$  & $878$  & $1794$ & $6090$ & $16947$ \\
		\JsReducedData         &                                   &                                                 & $46$   & $79$   & $135$  & $322$  & $787$   \\
		\JsWindowThreeData     &                                   &                                                 & $44$   & $56$   & $73$   & $109$  & $152$   \\
		\JsWindowFiftyData     &                                   &                                                 & $396$  & $578$  & $768$  & $1036$ & $1332$  \\

		\bottomrule
	\end{tabular}
	\\[.1cm]
	\label{table:data-stats}
\end{table*}

\JsOriginalData{} dataset comprises pre-files scraped from GitHub in their entirety. In the post-files, only changes relevant to the fix are retained, excluding irrelevant modifications like style fixes, refactoring, or new features. As highlighted in Section \ref{sec:approach-general}, \codereduce is designed to ignore such changes. To ensure a fair experimental setup in comparison to \codereduce{}, we take precautions to filter noisy changes in this dataset. Without this control, the noisy changes would lead to unfair comparisons in the metrics dependent on direct code comparison and also potentially harm the learning process.

\JsReducedData{} data contains the diffs from \JsOriginalData{} data passed through the \codereduce{} algorithm. This results in a much smaller-sized source code which, importantly, does not contain parts unrelated to the issue at hand.

\JsWindowFiftyData{}  data is a result of keeping $\pm50$ lines of code around the line of the detected issue and its corresponding fix. This type of data serves for experimenting with baseline models.

\JsWindowThreeData{}  data is a result of keeping $\pm3$ lines of code around the line of the detected issue and its corresponding fix. This type of data also serves for experimenting with baseline models.

Dataset statistics are provided in~Table~\ref{table:data-stats}. We also quantify the ratio between token lengths of the \JsOriginalData{} and \JsReducedData{} source code in Table~\ref{table:compression-level-per-category}.

\begin{table}
	\centering
	\caption{\codereduce{}'s compression ratio for $\{\text{\JsOriginalData{}}\mapsto\text{\JsReducedData}\}$.}
	\footnotesize
	\begin{tabular}[c]{@{\quad}l*{2}{>{\centering\arraybackslash}p{\widthof{XXXXXXXXXXX}}}@{\quad}}
		\toprule
		\multirow{2}{*}{Issue Category} & \multicolumn{2}{c}{Compression ratio (token size)}                    \\
		\cmidrule{2-3}
		                                & Train                                              & Test             \\
		\midrule
		\AST                            & $\times$ $40.67$                                   & $\times$ $57.44$ \\
		\Local                          & $\times$ $28.78$                                   & $\times$ $22.44$ \\
		\SecurityLocal                  & $\times$ $22.86$                                   & $\times$ $23.60$ \\
		\FileWide                       & $\times$ $19.83$                                   & $\times$ $10.96$ \\
		\SecurityFlow                   & $\times$ $19.46$                                   & $\times$ $20.80$ \\
		\bottomrule
	\end{tabular}
	\\[.2cm]
	\label{table:compression-level-per-category}
	\vspace{-.5cm}
\end{table}

\section{Experimental Evaluation}\label{sec:experiments}
We present an extensive evaluation of \tool{}.

\subsection{Train/Test Split}\label{sec:datasplit}

In many domains, the conventional approach involves randomly partitioning the dataset into training and test sets. However, this methodology, commonly employed in program repair research, is inherently unsuitable for the field of automated program repair. Randomly splitting the data may inadvertently lead to instances from the same repository or even the file being allocated to both the training and test sets, resulting in data leakage. Furthermore, the use of pretrained models introduces another potential source of data leakage. These models have been trained on extensive datasets that encompass open-source code, creating the possibility that samples in the fine-tuning test dataset were also present in the pretraining dataset of the underlying model. Remarkably, these two concerns have not received widespread attention in the context of Automated Program Repair.

We introduce an advanced train/test splitting that effectively addresses both of these issues. In addition to utilizing code with permissive licenses, we conducted data mining from open-source repositories governed by restrictive licenses. These restrictive licenses prohibit the utilization of code for learning but allow its use for testing. Consequently, we constructed our test set from these repositories. Note that this approach results in the utilization of entirely distinct sets of repositories for training and testing, eliminating the potential for data leakage stemming from shared repositories and files. Moreover, this approach effectively mitigates data leakage arising from the pretrained model's dataset. Given that the repositories covered by non-permissive licenses were not suitable for training, we can reasonably assume that they were not utilized by prior works either.

Finally, this allows us to have a larger set of test samples as opposed to traditional splits where one is usually concerned with losing training samples. Our test dataset contains \NumTestSamples{} samples and they were manually labelled in the same way as the training samples.

\subsection{Metrics}\label{sec:metrics}

For measuring models' performance, we utilize two metrics: functional correctness (\PassAtKMetric{}) and exact match to the target code (\ExactMatchAtKMetric{}). Both metrics are computed using ``@$k$'' flavor, that is, at different amounts of predictions generated by beam search~\cite{Graves2012SequenceTW} or nucleus sampling \cite{nucleussampling} depending on the model. To simplify the explanation, we consider the following notation: for input code $C$ with an issue $I$ instance detected on line $\ell$, the model generates $k$ outputs $\bm{p} \equiv \bm{p}(C, \ell, I) = (p_1, \ldots, p_k)$.

\PassAtKMetric{}: The code analysis engine is capable, for the given issue instance $(C, \ell, I)$ and a corresponding prediction $p$, to provide a predicate $\DoesFix(p, C, \ell, I)$, capturing if this prediction fixes the given issue instance. Note that line information is important, because there can be multiple instances of the same issue on several lines. In addition to that and following the common practice (e.g. see~\cite{TFix}), another predicate $\NoNewIssues(p|C)$ can check if $p$ introduced new issues elsewhere in the code compared to $C$. Using this information, the \PassAtKMetric\ checks if any of the $k$ predictions fixed the static analysis report \textit{and} not introduced any new issues, and takes the average:
\setlength{\arraycolsep}{0.0em}

\begin{equation}
	\footnotesize
	\begin{aligned}
		\text{\PassAtKMetric} =
		\frac{1}{|\bigl\{(C, \ell, I)\bigr\}|}
		\sum_{(C, \ell, I)}
		\max_{p \in \bm{p}} \bigl[\DoesFix(p, C, \ell, I) \\ \wedge \NoNewIssues(p|C)\bigr].
	\end{aligned}
\end{equation}

Naturally, $\DoesFix(\ldots)$ and $\NoNewIssues(\ldots)$ require that the prediction is parseable and syntactically correct code.

\ExactMatchAtKMetric{}: This metric relies on the existence of the labeled target fix $C'$ for the given issue instance $(C, \ell, I)$ and computes if any of the predictions equals the labeled target code $C'$:
\begin{equation}
	\footnotesize
	\text{\ExactMatchAtKMetric} = \frac{1}{|\bigl\{(C, C', \ell, I)\bigr\}|}\sum_{(C, C', \ell, I)} \max_{p \in \bm{p}} \bigl[ p = C' \bigr]
\end{equation}

Both metrics have their advantages and disadvantages. The quality of \PassAtKMetric{} depends on the robustness of the analysis engine. For example, a trivial definition of $\mathrm{DoesFix(\ldots)}$ as ``the issue is not detected anymore'' can be easily overfitted to~--- e.g., by deleting the whole code. While the code analysis engine used in our experiments is much more advanced, it is not free of producing false positive in certain corner cases. Defining a ``suitable'' metric for program repair proved to be hard. Reporting functional correctness and exact match has the advantage of relaying the trade-off between ``creativeness'' and ``canonicalization'' of fixes, with \PassAtKMetric{} and \ExactMatchAtKMetric{} being the upper- and lower-bound for some ``true'' metric.

\subsection{Training and Inference Configuration}

\textbf{Open-source models (StarCoderBase, T5, Mistral, Mixtral):}
We fine-tuned them on 16 Nvidia A100 GPUs for $60$ epochs with batch size of $4$ per device, gradient accumulation steps of $16$, learning rate of $10^{-5}$ with linear scheduler and warm-up ratio of $0.1$. We use DeepSpeed~\cite{deepspeed2020} with ZeRO-3 optimization~\cite{rajbhandari2020zero}. We use beam search and report various ``$@k$'' metrics. Specifically for Mixtral, we used QLoRA~\cite{QLora} to effectively fine-tune it as the model is relatively large.

\textbf{Models accessible via API (GPT-3.5, GPT-4):}
GPT-3.5~\cite{gpt35} and GPT-4~\cite{gpt4} allow querying via APIs. We employed the most recent releases, namely \emph{gpt-3.5-turbo-0613} and \emph{gpt-4-0613}, as of November 2023. We accessed the models via APIs, and ran the inference on our test set with few-shot learning.

\textbf{Few-shot examples choice:} for each of the samples requesting to fix an issue $I$, the few-shot training examples were chosen randomly from the training dataset for the same issue type. For each query, we provided 1 and 2 few-shot example(s) for GPT-3.5 and GPT-4, respectively. This choice was made based on the context window limitations of each model. Seeds were preserved to keep few-shot examples consistent across experiments.

\textbf{Prompt structure:} we used the best practices ~\cite{gptbestpractices} issued for GPT-3.5 and GPT-4 by OpenAI to set up the prompt and the conversations. For more in-depth details on the shape and structure of the prompts used, see Appendix~\ref{sec:appendix-llm-prompts}.

\textbf{Inference hyperparameters:} We set the maximum number of generated tokens to the maximum context size of the model and used temperature of $0.2$ for cases where we generate multiple predictions. While querying GPT-4 with full file in a zero-shot setting, the temperature was $0$. The instructions we provided in the system prompt (see Appendix~\ref{sec:appendix-llm-prompts}) were sufficient to configure the model's output.

\subsection{Comparison with Baselines}\label{sec:exp-codet5-window-3}

\begin{table*}[!htbp]
	\centering
	\tiny
	\renewcommand{\arraystretch}{1.2}
	\caption{Evaluation of \PassAtKMetric{} and \ExactMatchAtKMetric{} metrics for models that use \codereduce{} (marked as $^{\bm{\ddag}}$) against baselines of TFix~\cite{TFix}, and various large-context models. Model(s) corresponding to our approach is marked as $^{\bm{\ddag}}$. Refer to Section~\ref{sec:exp-codet5-window-3} for discussion. For examples of fixes produced by \tool{}, see Appendix~~\ref{sec:appendix_examples}.}
\begin{tabular}[c]{@{\quad}ll*{6}{>{\centering\arraybackslash}p{\widthof{\;$k=5$\;}}}@{\quad}}
		\toprule
		\mulro[.6]{2}{Issue Category} & \mulro[.6]{2}{Model}
		                              & \multicolumn{3}{c}{\PassAtKMetric{} ($\%$)}                                                        & \multicolumn{3}{c}{\ExactMatchAtKMetric{} ($\%$)}                                                                                                                               \\
		\cmidrule(l{1pt}r{2pt}){3-5} \cmidrule(l{1pt}r{2pt}){6-8}
		                              &                                                                                                    & $k = 1$                                           & $k=3$                      & $k = 5$                    & $k = 1$                    & $k=3$                      & $k = 5$ \\
		\midrule
		\mulro{10}{\AST}
		                              & \rlap{\textsc{Mixtral-8x7B}}\phantom{\textsc{Mixtral-8x7B}}\textsc{-\JsReducedData}$^{\bm{\ddag}}$

		                              & $81.01$                                                                                            & $88.96$                                           & $89.16$                    & $39.29$                    & $\bm{52.25}\rlap{$^\ast$}$ & $\bm{53.5}\rlap{$^\ast$}$            \\

		\cmidrule(l{1pt}r{1pt}){2-2}
		                              & \rlap{\textsc{StarCoder-7B}}\phantom{\textsc{Starcoder-7B}}\textsc{-\JsReducedData}$^{\bm{\ddag}}$

		                              & $\bm{70.05}$                                                                                       & $\bm{80.84}$                                      & $\bm{82.31}$               & $\bm{41.19}$               & $\bm{48.15}$               & $\bm{48.56}$                         \\
\bottomrule
	\end{tabular}
	\\[.2cm]
	\label{table:results-reduced-vs-window}
	\vspace{-.75cm}
\end{table*}

In the following main experiment, we aimed to compare bug-fixing performance of different models under different context extraction methods and learning setups, including models like GPT-4 ~-- thus forming strong baselines to test our approach. We use text-to-text transformers pre-trained on code or natural language (StarCoderBase~\cite{starcoder}, T5~\cite{t5}, Mixtral~\cite{Mixtral}) or for few-shot learning via API (GPT-3.5~\cite{gpt35}, GPT-4~\cite{gpt4}). Further, we utilize three datasets, \JsReducedData{} (represents our approach to context extraction) \JsWindowFiftyData{} (represents long context) and \JsWindowThreeData{} (represents a baseline, see e.g.~\cite{TFix}).
As a result, the following models were evaluated:

\textbf{StarCoderBase:} StarCoderBase model~\cite{starcoder}, fine-tuned by us on the \JsReducedData{} data \JsWindowFiftyData{}, allowing us to compare both approaches against each other in a fine-tuning setup.

\textbf{TFix~\cite{TFix} (also named \textsc{T5-\JsWindowThreeData} for clarity):} T5 model fine-tuned by us on the \JsWindowThreeData{} data, i.e.~only $3$ lines before/after the issue are taken as the input to the model. Thus, the model only utilizes linear span of context lines to be trained on.

\textbf{GPT-3.5 \& GPT-4:} Predictions for the \JsReducedData{}, \JsWindowFiftyData{} and \JsOriginalData{} are obtained via few-shot learning from the model.

\textbf{Mixtral-8x7B:} Predictions for the \JsReducedData{} are obtained via fine-tuning with QLoRA. Due to the model size and limited resources, we could not fine-tune this model on \JsWindowFiftyData{}, which would strengthen our results.

\textbf{Metrics on \JsReducedData{}:} During inference, after obtaining predictions, they are merged back into the full source code by means of the \mergeback{} algorithm, thus resulting in predicting the complete file. The metrics are then computed on the full files.

\textbf{Metrics on \JsWindowFiftyData{} \& \JsWindowThreeData{}:} During inference, after obtaining such ``windowed'' predictions, they are pasted back into the full source code, and then the metrics are computed on the full files.

Results are summarized in Table~\ref{table:results-reduced-vs-window}. We can observe that our approaches take advantage of ``compressing'' the context through reduction: they outperform the \JsWindowFiftyData{}, \JsOriginalData{} and \JsWindowThreeData{} baselines by a considerable margin w.r.t.~the \PassAtKMetric{} and \ExactMatchAtKMetric{}.

For the GPT model family, within each pair of \JsReducedData-\JsWindowFiftyData{} and \JsReducedData-\JsOriginalData{}, the former leads to better \PassAtKMetric. The gap is especially large for issue categories with complex data flow. For GPT-4 and \SecurityFlow{}, \codereduce{} improves $23\%$ over \JsWindowFiftyData{}, while improving $14\%$ over \JsOriginalData{}. We also see that \textsc{GPT-4-32K-\JsOriginalData{}-ZeroShot} yields not better, but competitive performance in some categories but querying the model in this way has a considerable latency. Hence, we evaluated it only for $k=1$. To be fair, we also evaluted \textsc{GPT-4-\JsReducedData{}-ZeroShot} with $k=1$ and our approach yields better results in all categories, especially for \SecurityFlow{}, \FileWide{} and \Local{}.

We note that there is a range of other bugfixing models such as SequenceR~\cite{sequencer}, CoCoNuT~\cite{encore} or Hoppity~\cite{hoppity} that were already shown to be noncompetitive to the T5 model used in TFix by prior works (\cite{visiontransformerrepair}, \cite{TFix}). We omit them from our evaluation as our approach already outperforms TFix by a large margin.
We also investigated the potential effects of model size between \textsc{1B} and \textsc{7B} and model architecture between \textsc{StarCoderBase} and \textsc{Mistral}~\cite{Mistral}. The results are summarized in Appendix~\ref{sec:appendix_size_arch}. In contrast to the findings observed by~\cite{TFix}, we conclude that increasing model size lead to significant and consistent advantages. Also, the pre-trained model and its architecture can make a significant difference. Finally, examples of the model predictions, fixing long-ranging issues that benefit from using \codereduce{}, can be found in Appendix~\ref{sec:appendix_examples}.

\subsection{Effects of \mergeback{}}

Compared to long context or window-based baselines, our approach using \JsReducedData{} contains more non-trivial steps, including \mergeback{}. In Table~\ref{table:results-mergeback-effect} we assess the effect of \mergeback{} on metrics. Note that this is possible because these \PassAtKMetric/\ExactMatchAtKMetric{} are computable also on reduced (before \mergeback) code, because one can validate if the fix removes a static analysis report on reduced code.

\begin{table*}
	\centering
	\caption{Influence of \mergeback{} on the metrics for \textsc{StarCoderBase-\JsReducedData}. Metrics are calculated on the full source code and reduced code, respectively.}
	\renewcommand{\arraystretch}{1.2}
	\scriptsize

	\begin{tabular}[c]{@{\quad}ll*{4}{>{\centering\arraybackslash}p{\widthof{\;$00.000$\;}}}@{\quad}}
		\toprule
		\mulro[.6]{2}{Issue Category} & \mulro[.6]{2}{Step} & \multicolumn{2}{c}{\PassAtKMetric{} ($\%$)} & \multicolumn{2}{c}{\ExactMatchAtKMetric{} ($\%$)}                     \\
		\cmidrule(l{1pt}r{2pt}){3-4} \cmidrule(l{2pt}r{1pt}){5-6}
		                              &                     & $k=1$                                       & $k=3$                                             & $k=1$   & $k=3$   \\

		\midrule
		\mulro{2}\AST{}               & after \mergeback{}  & $70.05$                                     & $80.84$                                           & $41.19$ & $48.15$ \\
		                              & before \mergeback{} & $78.97$                                     & $88.63$                                           & $41.19$ & $48.15$ \\

		\midrule
		\mulro{2}\Local{}             & after \mergeback{}  & $82.75$                                     & $89.72$                                           & $34.23$ & $42.02$ \\
		                              & before \mergeback{} & $81.60$                                     & $90.85$                                           & $34.29$ & $42.02$ \\

		\midrule
		\mulro{2}\FileWide{}          & after \mergeback{}  & $77.38$                                     & $80.16$                                           & $44.59$ & $54.50$ \\
		                              & before \mergeback{} & $87.62$                                     & $92.06$                                           & $44.59$ & $54.50$ \\

		\midrule
		\mulro{2}\SecurityLocal{}     & after \mergeback{}  & $74.68$                                     & $89.08$                                           & $20.18$ & $29.53$ \\
		                              & before \mergeback{} & $79.20$                                     & $93.21$                                           & $20.18$ & $29.53$ \\

		\midrule
		\mulro{2}\SecurityFlow{}      & after \mergeback{}  & $49.15$                                     & $65.90$                                           & $10.63$ & $13.49$ \\
		                              & before \mergeback{} & $52.40$                                     & $69.26$                                           & $10.63$ & $13.49$ \\

		\bottomrule
	\end{tabular}

	\label{table:results-mergeback-effect}
\end{table*}

One can observe that the \mergeback{} algorithm is not perfect in the sense that, compared to evaluating reduced predictions, it introduced a certain drop of \PassAtKMetric{}, while \ExactMatchAtKMetric{} does not suffer. One of the reasons is that as the model performs more ``creative'' fixes, not all of them are possible to integrate back in the original large code. It should be noted that the performance of our approach can be further increased by improving the \mergeback{} algorithm. We leave tightening this gap as future work.

\subsection{Latency of the end-to-end bugfixing system}


We performed an experiment to compare the algorithm of \codereduce{} (Algorithm \ref{algo:codereduce}) with a vanilla HDD algorithm~\cite{DBLP:HDD}. To avoid differences caused by the machine and analyzer speeds, we evaluate the number of calls to the analyzer made by code reduction algorithms. For this evaluation, we took $1024$ programs from our training dataset and measured the code reduction speed using both settings. The number of program analyzer calls executed by Algorithm \ref{algo:codereduce} averages to $9.44$ and has geometric mean of $4.99$, whereas the number of calls to the program analyzer the HDD algorithm performs averages to $11.52$ and has geometric mean of $7.25$. When taking into account geometric means that are useful for computing ratios, it leads to a ratio of $0.68$ steps for the improved algorithm versus HDD or $30\%$ reduction. When comparing averages, the reduction is $20\%$, due to several outlier samples. The most complex example in our data required $181$ calls to the analyzer.

We performed the evaluation on an AWS \texttt{c5ad.8xlarge} instance, running Ubuntu 22.04. In terms of latency, each call to the underlying static analyzer takes around $70$ to $80ms$. This means the entire \codereduce{} algorithm takes less than $1$ second on average. Querying \textsc{StarCoderBase-7B} takes on average $605ms$ per prediction with an A100 GPU. Taking into account the need to check the solution after \mergeback{} and generating several candidates from the language model, the proposed bugfixing system performs predictions within a few seconds and is usable in an IDE setting.

\section{Related Work}\label{sec:related-work}



Automated program repair (APR) has been studied both from traditional software engineering and machine learning perspectives \cite{repair-living-review, DBLP:journals/tse/GazzolaMM19}. Traditional approaches typically require test cases, crash reports or other specifications. Given a program violating the expected behaviour, APR systems output a correct program fulfilling the expected behaviour, usually by performing a smart search over the space of program modifications. However, most of these works end up being unpractical due to the enormous search space, continuous recompilation of candidate patches, and re-execution of the tests \cite{towardspracticalprogramrepair}. Some works try to synthesize the output by solving symbolic constraints \cite{angelix, semfix, nopol} or applying learned edit patterns on abstract syntax trees \cite{DBLP:conf/sigsoft/LongAR17, getafix, Sorald}. Both approaches typically suffer from not generalizing as the behaviour is specific to individual programs \cite{DBLP:conf/sigsoft/SmithBGB15} or suffer from overfitting \cite{DBLP:journals/jss/YeMDM21}. Our work significantly differs from prior APR approaches both technically and usability. \tool{} generalizes across different programs by learning from a large dataset of diverse bugs and fixes, scales to larger edits and generates fixes practically instantly in comparison to APR works.

Recent years have seen an increase in learning-based tools. A major hurdle for these systems is the lack of training data. Several works attack this problem by generating synthetic fixes for specific types of bugs like incorrect operators, variable misuses \cite{DBLP:journals/pacmpl/PradelS18, semanticcoderepairusingneurosymbolic}. Applying various models such as LSTMs \cite{DBLP:conf/iclr/VasicKMBS19} and Transformers \cite{DBLP:conf/iclr/HellendoornSSMB20} increase the accuracy on the synthetic datasets, but these tools still produce mostly false positives on real-world bug distributions \cite{DBLP:conf/nips/AllamanisJB21}. As a result, these systems have limited practical applicability and the problem of the distribution shift by learning from synthetic data remains an open research question \cite{DBLP:conf/icml/HeBV22}. Other recent works use large language models to transform programs with errors into programs without errors \cite{deepfix, TFix}. Another line of work~\cite{Synchromesh, RING} proposes to replace fine-tuning the model with selecting several training data samples similar to the query and using them as a prompt in a few-shot prediction of a larger pretrained model for code~\cite{codex}. All of these works are limited to relatively small programs that fit into the context size of the model they use.


\section{Conclusion}\label{sec:conclusion}

We presented a new learning-based system, \tool{}, for automatically fixing non-trivial coding errors and security vulnerabilities. The key insight is to rely on a code reduction mechanism that incorporates program analysis into the machine learning pipeline. \tool{} learns to fix non-trivial coding errors accurately as \codereduce{} extracts the essential information needed for the fix, allowing the learning process to avoid having to uncover arbitrarily long-range dependencies and focusing it on attending to the crucial parts of code.

Our analysis of open-source commits reveals insights about the difficulty of collecting large amounts of data for program repair. As a solution, we built a high-quality dataset bugs and fixes extracted from millions of real-world \github{} commits.

We evaluated our approach on multiple models - StarCoderBase, Mixtral, T5, GPT-3.5 and GPT-4 and demonstrated that directly using a model which tries to learn the complex long-range dependencies needed to produce a correct fix, performs generally worse than using a context extracted by \codereduce{}. Overall, \tool{} outperformed the previous state-of-the-art baseline TFix and helped multiple models in different setups to perform better as it utilizes program analysis to deal with long-range dependencies and data flows, greatly simplifying the attention learning task.

Although this paper's focus is on program repair, we believe our work is an important step in incorporating program analysis into machine learning and clearly highlights the need for code analysis, even when using powerful LLMs.

\bibliographystyle{ACM-Reference-Format}
\bibliography{bibliography/references}

\appendix
\onecolumn
\section{Rule Scope}\label{sec:rule-scope}
\begin{longtable}[c]{@{\quad}lc*{2}{>{\centering\arraybackslash}p{\widthof{XXX}}}}
	\caption{Issues reported by the code analysis, and their respective issue categories. We are actively labeling and enlarging our dataset. \label{table:rule-groups}} \\
	\toprule
	\mulro[.6]{2}{Issue Identifier}                                      & \mulro[.8]{2}{\shortstack{Issue                                                               \\Category}} & \multicolumn{2}{c}{\# Datapoints}                       \\
	\cmidrule{3-4}
	                                                                     &                                 & Train
	                                                                     & Test                                                                                          \\
	\midrule
	\endfirsthead
	\multicolumn{4}{c}{\textit{(table continued)}}                                                                                                                       \\
	\midrule
	\mulro[.6]{2}{Issue Identifier}                                      & \mulro[.8]{2}{\shortstack{Issue                                                               \\Category}} & \multicolumn{2}{c}{\# Datapoints} \\

	\cmidrule{3-4}
	                                                                     &                                 & Train
	                                                                     & Test                                                                                          \\
	\midrule
	\endhead
	\midrule
	\multicolumn{4}{c}{\textit{(table continued)}}                                                                                                                       \\
	\endfoot
	\bottomrule
	\endlastfoot

	AmbiguousConditional                                                 & \AST                            & 150         & 58                                            \\
	AngularSceProviderDisabled                                           & \AST                            & 2           & 1                                             \\
	ArgumentsAsParameter                                                 & \AST                            & 2           & 3                                             \\
	ArrayConstructor                                                     & \AST                            & 88          & 76                                            \\
	AssignmentWithSameVarOnLeftAndRight                                  & \AST                            & 8           & 6                                             \\
	BooleanObjectCreation                                                & \AST                            & 31          & 7                                             \\
	CommaOrSwitch                                                        & \AST                            & 18          & 4                                             \\
	ConditionWithAssign                                                  & \AST                            & 40          & 15                                            \\
\end{longtable}

\clearpage{}
\section{Prompts for GPT-3.5 \& GPT-4}\label{sec:appendix-llm-prompts}

Let \PromptRule{} and \PromptMessage{} denote the name and the description of the issue reported by the static analyzer in a code snippet. We will query the model to generate a fix for this code snippet and denote it by \PromptSource{}. Let $f$ be the number of few-shot examples provided in the prompt and the pair of code snippets (\FewShotSource{}, \FewShotTarget{}) denote the $ith$ example fix for the issue \PromptRule{} in the prompt.

GPT-3.5 and GPT-4 were designed to make conversations and we followed the best practices~\cite{gptbestpractices} shared by OpenAI to build the initial conversation. A conversation can contain three different roles, namely system, user and assistant. It is advised to start the conversation with a system content where one defines the role of the assistant (AI model) and gives instructions on the desired output structure. After that, we provide the few-shot examples as a conversation between the user and the assistant. The conversation is finished by the user providing the vulnerable code \PromptSource{} so that the next turn belongs to the assistant. The asistant completes the conversation by generating the fix to the last user query. Precisely, the following prompt is fed into GPT models. (The last sentence of the system prompt was not provided when the full file was fed into the model.)

\begin{align*}
	 & {\texttt{\scriptsize\PromptSystem{}
	\parbox{.4\textwidth}{Assistant is a code assistant designed to fix issues in given code snippets. Instructions: Do not generate additional text or code. Output only the fixed code snippet. Do not generate explanations, comments, notes. Note that the code we provide is incomplete, it is intentionally reduced to a smaller snippet, do not try to complete it in anyway. Leave evertything as it is and just apply the changes related to the fix.}}} \\
	 & {\parbox{.4\textwidth}{\texttt{\scriptsize\PromptUser{} Generate the fixed code for the bug \PromptRule{} with the error message \PromptMessage{}. \FewShotSourceOne{}}}}                                                                                                                                                                                                                                                                                  \\
	 & {\parbox{.4\textwidth}{\texttt{\scriptsize\PromptAssistant{} \FewShotTargetOne{}}}}                                                                                                                                                                                                                                                                                                                                                                        \\
	 & {\parbox{.4\textwidth}{\texttt{...}}}                                                                                                                                                                                                                                                                                                                                                                                                                      \\
	 & {\parbox{.4\textwidth}{\texttt{\scriptsize\PromptUser{} Generate the fixed code for the bug \PromptRule{} with the error message \PromptMessage{}. \FewShotSourceLast{}}}}                                                                                                                                                                                                                                                                                 \\
	 & {\parbox{.4\textwidth}{\texttt{\scriptsize\PromptAssistant{} \FewShotTargetLast{}}}}                                                                                                                                                                                                                                                                                                                                                                       \\
	 & {\parbox{.4\textwidth}{\texttt{\scriptsize\PromptUser{} Generate the fixed code for the bug \PromptRule{} with the error message \PromptMessage{}. \PromptSource{}}}}                                                                                                                                                                                                                                                                                      \\
\end{align*}

\clearpage{}
\section{Examples Fixes}\label{sec:appendix_examples}

\begin{figure*}[h]
	\vspace{-2mm}
	\centering
	\begin{subfigure}{.495\columnwidth}
		\begin{lstlisting}[language=jsappdx, title={(a) Input: vulnerable pre-version}, captionpos=t, numberstyle=\linecolors{white}{0}{0}]
const express = require('express')
const router = express.Router()

const dbConfig = require('../db/dbConfig')
const mysql = require('mysql')
const pool = mysql.createPool(dbConfig.mysql)

let responseJSON = function (res, ret) {
  if (typeof ret === 'undefined') {
    res.json({
      code: '-200', msg: 'failed operation'
    })
  } else {
    res.json(ret)
  }
}

// ... <REDACTED>

router.get('/api', (req, res, next) => {
	// ... <REDACTED>
  var obj = { name: 'huangming', age: 1 }
  res.json(obj)
})

// ... <REDACTED>

router.get('/postAdvice', (req, res, next) => {
  res.header('Access-Control-Allow-Origin', '*')
  res.header('Access-Control-Allow-Methods', 'PUT, GET, POST, DELETE, OPTIONS')
  res.header('Access-Control-Allow-Headers', 'X-Requested-With')
  res.header('Access-Control-Allow-Headers', 'Content-Type')
  let ip = req.headers['x-forwarded-for'] ||
    req.connection.remoteAddress ||
    req.socket.remoteAddress ||
    (req.connection.socket ? req.connection.socket.remoteAddress : null)
  pool.getConnection((err, connection) => {
    let param = req.query
(*\HL{myred}*)  	let sql = 'INSERT INTO w_advice (username,advice,email,ip) \
(*\HL{myred}*)  							VALUES ("' \
(*\HL{myred}*)  							+ param.username + '","' + param.advice \
(*\HL{myred}*)  							+ '","' + param.email + '","' + ip + '")' 

(*\HL{myred}\RoundedFrameLine{}*)  	connection.query(sql, (err, result) => {
(*\HL{myred}*)
(*\HL{myred}*)
(*\HL{myred}*)
      responseJSON(res, result)
    })
    connection.release()
  })
})
    \end{lstlisting}
	\end{subfigure}\hfill
	\begin{subfigure}{.495\columnwidth}
		\begin{lstlisting}[language=jsappdx, title={(b) Output: non-vulnerable full file}, captionpos=t, numberstyle=\linecolors{white}{0}{0}]		
const express = require('express')
const router = express.Router()

const dbConfig = require('../db/dbConfig')
const mysql = require('mysql')
const pool = mysql.createPool(dbConfig.mysql)

let responseJSON = function (res, ret) {
  if (typeof ret === 'undefined') {
    res.json({
      code: '-200', msg: 'failed operation'
    })
  } else {
    res.json(ret)
  }
}

// ... <REDACTED>

router.get('/api', (req, res, next) => {
	// ... <REDACTED>
  var obj = { name: 'huangming', age: 1 }
  res.json(obj)
})

// ... <REDACTED>

router.get('/postAdvice', (req, res, next) => {
  res.header('Access-Control-Allow-Origin', '*')
  res.header('Access-Control-Allow-Methods', 'PUT, GET, POST, DELETE, OPTIONS')
  res.header('Access-Control-Allow-Headers', 'X-Requested-With')
  res.header('Access-Control-Allow-Headers', 'Content-Type')
  let ip = req.headers['x-forwarded-for'] ||
    req.connection.remoteAddress ||
    req.socket.remoteAddress ||
    (req.connection.socket ? req.connection.socket.remoteAddress : null)
  pool.getConnection((err, connection) => {
    let param = req.query
(*\HL{mygreen}*)		let sql = 'INSERT INTO w_advice (username,advice,email,ip) \
(*\HL{mygreen}*)								VALUES (?, ?, ?, ?)'
(*\HL{mygreen}*)								
(*\HL{mygreen}*)

(*\HL{mygreen}*)  	connection.query(
(*\HL{mygreen}*)										sql, 
(*\HL{mygreen}*)										[param.username, param.advice, param.email, ip], 
(*\HL{mygreen}*)										(err, result) => {
      responseJSON(res, result)
    })
    connection.release()
  })
})
	  \end{lstlisting}
	\end{subfigure}

	\caption{\tool{} fixes \textit{Sql Injection}, one of the most common and critical vulnerabilities.
	}
\end{figure*}

\begin{figure*}
	\vspace{-2mm}
	\centering
	\begin{subfigure}{.495\columnwidth}
		\begin{lstlisting}[language=jsappdx, title={(a) Input: vulnerable pre-version}, captionpos=t, numberstyle=\linecolors{white}{0}{0}]
// ... <REDACTED>

import fs = require('fs')
import { type Request, type Response, type NextFunction } from 'express'

// ... <REDACTED>

module.exports = function profileImageUrlUpload () {
  return (req: Request, res: Response, next: NextFunction) => {
    if (req.body.imageUrl !== undefined) {
(*\HL{myred}*)  		const url = req.body.imageUrl
(*\RoundedFrameLine{}*)      if (url.match(/(.)*solve\/challenges\/server-side(.)*/) !== null) 
				req.app.locals.abused_ssrf_bug = true
      const loggedInUser = security.authenticatedUsers.get(req.cookies.token)
			// ... <REDACTED>
		  if (loggedInUser) {
				const imageRequest = request
					.get(url)
          .on('error', function (err: unknown) {
						// ... <REDACTED>
          })
          .on('response', function (res: Response) {
						// ... <REDACTED>
					} 
					// ... <REDACTED>
			}
		} 
	}
}

    \end{lstlisting}
	\end{subfigure}\hfill
	\begin{subfigure}{.495\columnwidth}
		\begin{lstlisting}[language=jsappdx, title={(b) Output: non-vulnerable full file}, captionpos=t, numberstyle=\linecolors{white}{0}{0}]		
// ... <REDACTED>

import fs = require('fs')
import { type Request, type Response, type NextFunction } from 'express'

// ... <REDACTED>

module.exports = function profileImageUrlUpload () {
  return (req: Request, res: Response, next: NextFunction) => {
    if (req.body.imageUrl !== undefined) {
(*\HL{mygreen}*)  		const url = typeof req.body.imageUrl === 'string' ? req.body.imageUrl : ''
      if (url.match(/(.)*solve\/challenges\/server-side(.)*/) !== null) 
				req.app.locals.abused_ssrf_bug = true
      const loggedInUser = security.authenticatedUsers.get(req.cookies.token)
			// ... <REDACTED>
		  if (loggedInUser) {
				const imageRequest = request
					.get(url)
          .on('error', function (err: unknown) {
						// ... <REDACTED>
          })
          .on('response', function (res: Response) {
						// ... <REDACTED>
					} 
					// ... <REDACTED>
			}
		} 
	}
}
	  \end{lstlisting}
	\end{subfigure}

	\caption{\tool{} fixes a \textit{HTTPSourceWithUncheckedType} vulnerability inside \textit{Juice-Shop}, one of the intentionally vulnerable benchmark repositories.
	}
\end{figure*}

\begin{figure*}
	\vspace{-2mm}
	\centering
	\begin{subfigure}{.495\columnwidth}
		\begin{lstlisting}[language=jsappdx, title={(a) Input: vulnerable pre-version}, captionpos=t, numberstyle=\linecolors{white}{0}{0}]
// ... <REDACTED>
const finale = require('finale-rest')
(*\HL{myemptygray}*)
const express = require('express')
const compression = require('compression')
// ... <REDACTED>
const models = require('./models')
const datacreator = require('./data/datacreator')
(*\RoundedFrameLine{}*)const app = express()
(*\HL{myemptygray}*)
// ... <REDACTED>
const collectDurationPromise = (name, func) => {
  return async (...args) => {
    const end = startupGauge.startTimer({ task: name })
    const res = await func(...args)
    end()
    return res
  }
}

// ... <REDACTED>

/* Sets view engine to hbs */
app.set('view engine', 'hbs')

restoreOverwrittenFilesWithOriginals().then(() => {
	// ... <REDACTED>
	app.use(errorhandler())
}).catch((err) => {
	// ... <REDACTED>
})
	
// ... <REDACTED>

    \end{lstlisting}
	\end{subfigure}\hfill
	\begin{subfigure}{.495\columnwidth}
		\begin{lstlisting}[language=jsappdx, title={(b) Output: non-vulnerable full file}, captionpos=t, numberstyle=\linecolors{white}{0}{0}]		
// ... <REDACTED>
const finale = require('finale-rest')
(*\HL{mygreen}*)import csrf = require('csurf')
const express = require('express')
const compression = require('compression')
// ... <REDACTED>
const models = require('./models')
const datacreator = require('./data/datacreator')
const app = express()
(*\HL{mygreen}*)app.use(csrf({ cookie: true }))
// ... <REDACTED>
const collectDurationPromise = (name, func) => {
  return async (...args) => {
    const end = startupGauge.startTimer({ task: name })
    const res = await func(...args)
    end()
    return res
  }
}

// ... <REDACTED>

/* Sets view engine to hbs */
app.set('view engine', 'hbs')

restoreOverwrittenFilesWithOriginals().then(() => {
	// ... <REDACTED>
	app.use(errorhandler())
}).catch((err) => {
	// ... <REDACTED>
})
	
// ... <REDACTED>
	  \end{lstlisting}
	\end{subfigure}

	%
	%
	%
	%
	\caption{\tool{} fixes a \textit{UseCsrfForExpress} vulnerability inside \textit{Juice-Shop}. Note that the import statement for \textit{express}, the definition of the app and its usage can be arbitarly far away from each other. \tool{} brings them all in the same range and achieves to modify several places in the file without any issue.
	}
\end{figure*}

\begin{figure*}
	\vspace{-2mm}
	\centering
	\begin{subfigure}{.495\columnwidth}
		\begin{lstlisting}[language=jsappdx, title={(a) Input: vulnerable pre-version}, captionpos=t, numberstyle=\linecolors{white}{0}{0}]
var router = require('express').Router()
var vulnDict = require('../config/vulns')
var authHandler = require('../core/authHandler')

(*\HL{myemptygray}*)
module.exports = function (passport) {
	router.get('/', authHandler.isAuthenticated, function (req, res) {
		res.redirect('/learn')
	})

	router.get('/login', authHandler.isNotAuthenticated, function (req, res) {
		res.render('login')
	})

	// ... <REDACTED>

	router.get('/learn', authHandler.isAuthenticated, function (req, res) {
		res.render('learn',{vulnerabilities:vulnDict})
	})

	router.get('/logout', function (req, res) {
		req.logout();
		res.redirect('/');
	})

	// ... <REDACTED>

	(*\HL{myemptygray}*)
	(*\HL{myemptygray}*)
	(*\HL{myemptygray}*)
	(*\HL{myemptygray}*)

	(*\RoundedFrameLine{}\HL{myred}*)router.post('/login', passport.authenticate('login', {
		successRedirect: '/learn',
		failureRedirect: '/login',
		failureFlash: true
	}))

	router.post('/register', passport.authenticate('signup', {
		successRedirect: '/learn',
		failureRedirect: '/register',
		failureFlash: true
	}))

	// ... <REDACTED>

	router.post('/resetpw', authHandler.resetPwSubmit)

	return router
}

    \end{lstlisting}
	\end{subfigure}\hfill
	\begin{subfigure}{.495\columnwidth}
		\begin{lstlisting}[language=jsappdx, title={(b) Output: non-vulnerable full file}, captionpos=t, numberstyle=\linecolors{white}{0}{0}]		
var router = require('express').Router()
var vulnDict = require('../config/vulns')
var authHandler = require('../core/authHandler')

(*\HL{mygreen}*)const rateLimit = require('express-rate-limit')
module.exports = function (passport) {
	router.get('/', authHandler.isAuthenticated, function (req, res) {
		res.redirect('/learn')
	})

	router.get('/login', authHandler.isNotAuthenticated, function (req, res) {
		res.render('login')
	})

	// ... <REDACTED>

	router.get('/learn', authHandler.isAuthenticated, function (req, res) {
		res.render('learn',{vulnerabilities:vulnDict})
	})

	router.get('/logout', function (req, res) {
		req.logout();
		res.redirect('/');
	})

	// ... <REDACTED>

	(*\HL{mygreen}*)const loginLimiter = rateLimit({
	(*\HL{mygreen}*)  windowMs: 15 * 60 * 1000, // 15 minutes
	(*\HL{mygreen}*)  max: 5
	(*\HL{mygreen}*)})

	(*\HL{mygreen}*)router.post('/login', loginLimiter, passport.authenticate('login', {
		successRedirect: '/learn',
		failureRedirect: '/login',
		failureFlash: true
	}))

	router.post('/register', passport.authenticate('signup', {
		successRedirect: '/learn',
		failureRedirect: '/register',
		failureFlash: true
	}))

	// ... <REDACTED>

	router.post('/resetpw', authHandler.resetPwSubmit)

	return router
}

	  \end{lstlisting}
	\end{subfigure}

	%
	%
	%
	%
	\caption{\tool{} fixes a \textit{NoRateLimiting} vulnerability inside \textit{appsecco/dvna}, one of the intentionally vulnerable benchmark repositories. This is a hard to fix vulnerability because the fix requires changes in 3 different locations of the file and ssome of those changes involve multiple lines.
	}
\end{figure*}

\begin{figure*}
	\vspace{-2mm}
	\centering
	\begin{subfigure}{.495\columnwidth}
		\begin{lstlisting}[language=jsappdx, title={(a) Input: vulnerable pre-version}, captionpos=t, numberstyle=\linecolors{white}{0}{0}]
// ... <REDACTED>
const exec = require('child_process').exec;
var mathjs = require('mathjs')
var libxmljs = require("libxmljs");
var serialize = require("node-serialize")
const Op = db.Sequelize.Op

// ... <REDACTED>

module.exports.ping = function (req, res) {
(*\RoundedFrameLine{}\HL{myred}*)  	exec('ping -c 2 ' + req.body.address, function (err, stdout, stderr) {
		output = stdout + stderr
		res.render('app/ping', {
			output: output
		})
	})
}

// ... <REDACTED>

module.exports.listProducts = function (req, res) {
	db.Product.findAll().then(products => {
		output = {
			products: products
		}
		res.render('app/products', {
			output: output
		})
	})
}

// ... <REDACTED>

module.exports.listUsersAPI = function (req, res) {
	db.User.findAll({}).then(users => {
		res.status(200).json({
			success: true,
			users: users
		})
	})
}
// ... <REDACTED>

    \end{lstlisting}
	\end{subfigure}\hfill
	\begin{subfigure}{.495\columnwidth}
		\begin{lstlisting}[language=jsappdx, title={(b) Output: non-vulnerable full file}, captionpos=t, numberstyle=\linecolors{white}{0}{0}]		
// ... <REDACTED>
const exec = require('child_process').exec;
var mathjs = require('mathjs')
var libxmljs = require("libxmljs");
var serialize = require("node-serialize")
const Op = db.Sequelize.Op

// ... <REDACTED>

module.exports.ping = function (req, res) {
(*\HL{mygreen}*)  	execFile('ping', ['-c', '2', req.body.address], function (err, stdout, stderr) {  	
		output = stdout + stderr
		res.render('app/ping', {
			output: output
		})
	})
}

// ... <REDACTED>

module.exports.listProducts = function (req, res) {
	db.Product.findAll().then(products => {
		output = {
			products: products
		}
		res.render('app/products', {
			output: output
		})
	})
}

// ... <REDACTED>

module.exports.listUsersAPI = function (req, res) {
	db.User.findAll({}).then(users => {
		res.status(200).json({
			success: true,
			users: users
		})
	})
}
// ... <REDACTED>
	  \end{lstlisting}
	\end{subfigure}

	%
	%
	\caption{\tool{} fixes a \textit{CommandInjection} vulnerability inside \textit{appsecco/dvna}, one of the intentionally vulnerable benchmark repositories. Note how \tool{} keeps the required import during reduction and the significant compression rate.
	}
\end{figure*}

\begin{figure*}
	\vspace{-2mm}
	\centering
	\begin{subfigure}{.495\columnwidth}
		\begin{lstlisting}[language=jsappdx, title={(a) Input: vulnerable pre-version}, captionpos=t, numberstyle=\linecolors{white}{0}{0}]
var express = require('express')
var bodyParser = require('body-parser')
var passport = require('passport')
var session = require('express-session')
var ejs = require('ejs')
var morgan = require('morgan')
const fileUpload = require('express-fileupload');
var config = require('./config/server')

//Initialize Express
(*\HL{myemptygray}*)
(*\RoundedFrameLine{}*)var app = express()
require('./core/passport')(passport)
app.use(express.static('public'))
app.set('view engine','ejs')
app.use(morgan('tiny'))
app.use(bodyParser.urlencoded({ extended: false }))
app.use(fileUpload());

// Enable for Reverse proxy support
// app.set('trust proxy', 1) 

// Intialize Session
app.use(session({
  secret: 'keyboard cat',
  resave: true,
  saveUninitialized: true,
  cookie: { secure: false }
}))

// Routing
app.use('/app',require('./routes/app')())
(*\HL{myemptygray}*)
app.use('/',require('./routes/main')(passport))

// Start Server
app.listen(config.port, config.listen)

    \end{lstlisting}
	\end{subfigure}\hfill
	\begin{subfigure}{.495\columnwidth}
		\begin{lstlisting}[language=jsappdx, title={(b) Output: non-vulnerable full file}, captionpos=t, numberstyle=\linecolors{white}{0}{0}]		
var express = require('express')
var bodyParser = require('body-parser')
var passport = require('passport')
var session = require('express-session')
var ejs = require('ejs')
var morgan = require('morgan')
const fileUpload = require('express-fileupload');
var config = require('./config/server')

//Initialize Express
(*\HL{mygreen}*)var helmet = require('helmet')
var app = express()
require('./core/passport')(passport)
app.use(express.static('public'))
app.set('view engine','ejs')
app.use(morgan('tiny'))
app.use(bodyParser.urlencoded({ extended: false }))
app.use(fileUpload());

// Enable for Reverse proxy support
// app.set('trust proxy', 1) 

// Intialize Session
app.use(session({
  secret: 'keyboard cat',
  resave: true,
  saveUninitialized: true,
  cookie: { secure: false }
}))

// Routing
app.use('/app',require('./routes/app')())
(*\HL{mygreen}*)app.use(helmet())
app.use('/',require('./routes/main')(passport))

// Start Server
app.listen(config.port, config.listen)
	  \end{lstlisting}
	\end{subfigure}

	%
	%
	%
	\caption{\tool{} fixes a \textit{UseHelmetForExpress} vulnerability inside \textit{appsecco/dvna}. The fix is seemingly simple as one can add \textit{helmet} with a single line. However, without adding the right import statement, the code will be broken. A great bug-fixing tool must apply imports correctly. This makes even the seemingly simple fix patterns much harder as the import statements and their usages can be arbitarly far away from each other. The rule \textit{UseHelmetForExpress} belongs to the category \textit{SecurityLocal} but it still requires changes in several different places of the file, just like other "\textit{Local}" rules.
	}
\end{figure*}

\begin{figure*}
	\vspace{-2mm}
	\centering
	\begin{subfigure}{.495\columnwidth}
		\begin{lstlisting}[language=jsappdx, title={(a) Input: vulnerable pre-version}, captionpos=t, numberstyle=\linecolors{white}{0}{0}]
'user strcit';
const config = require('./../../config')
var jwt = require("jsonwebtoken");
const { user } = require('../../orm');

module.exports = (app,db) => {
	app.post('/v1/user/token', (req,res) =>{
		// ... <REDACTED>
  });
  app.post('/v1/user/login', (req,res) =>{
		// ... REDACTED
  	const user = db.user.findAll({
    	where: {
      	email: userEmail
      }}).then(user => {
      })
		// ... REDACTED
  });
	// ... REDACTED
  app.put('/v1/admin/promote/:id', (req,res) =>{
  	const userId = req.params.id;
  	const user = db.user.update({role:'admin'}, {
    	where: {
      	id : userId
      }}
    )
    .then((user)=>{
(*\RoundedFrameLine{}\HL{myred}*)  		res.send(user)
		})
  });
  app.post('/v1/user/:id/validate-otp', (req,res) =>{
		// ... REDACTED
  	const user = db.user.findOne({
    	where: {
      	id: userId
      }}).then(user => {
				// ... REDACTED
      })
		// ... REDACTED
	});
};
    \end{lstlisting}
	\end{subfigure}\hfill
	\begin{subfigure}{.495\columnwidth}
		\begin{lstlisting}[language=jsappdx, title={(b) Output: non-vulnerable full file}, captionpos=t, numberstyle=\linecolors{white}{0}{0}]		
'user strcit';
const config = require('./../../config')
var jwt = require("jsonwebtoken");
const { user } = require('../../orm');

module.exports = (app,db) => {
	app.post('/v1/user/token', (req,res) =>{
		// ... <REDACTED>
  });
  app.post('/v1/user/login', (req,res) =>{
		// ... REDACTED
  	const user = db.user.findAll({
    	where: {
      	email: userEmail
      }}).then(user => {
      })
		// ... REDACTED
  });
	// ... REDACTED
  app.put('/v1/admin/promote/:id', (req,res) =>{
  	const userId = req.params.id;
  	const user = db.user.update({role:'admin'}, {
    	where: {
      	id : userId
      }}
    )
    .then((user)=>{
(*\HL{mygreen}*)  		res.status(200).json(user)
		})
  });
  app.post('/v1/user/:id/validate-otp', (req,res) =>{
		// ... REDACTED
  	const user = db.user.findOne({
    	where: {
      	id: userId
      }}).then(user => {
				// ... REDACTED
      })
		// ... REDACTED
	});
};
	  \end{lstlisting}
	\end{subfigure}

	\caption{\tool{} fixes a \textit{XSS} vulnerability inside \textit{SirAppSec/vuln-node.js-express.js-app}, one of the intentionally vulnerable benchmark repositories.
	}
\end{figure*}

\begin{figure*}
	\vspace{-2mm}
	\centering
	\begin{subfigure}{.495\columnwidth}
		\begin{lstlisting}[language=jsappdx, title={(a) Input: vulnerable pre-version}, captionpos=t, numberstyle=\linecolors{white}{0}{0}]
var requirejs = require('requirejs');
var config = requirejs('./config');

// ... <REDACTED>
    
var express = require('express'),
(*\HL{myred}*)    http = require('http'),
    path = require('path');

// ... <REDACTED>
    
var rootDir = path.join(__dirname, '..');

var app = express();

app.configure(function(){
  app.set('port', process.env.PORT || 3000);
  app.use(express.logger('dev'));
  // ... <REDACTED>
});

app.configure('development', function(){
  app.use('/app', 
    express.static(path.join(rootDir, 'app')));
  // ... <REDACTED>
});

(*\RoundedFrameLine\HL{myred}*)var server = http.createServer(app)
.listen(app.get('port'), function(){
  console.log("<REDACTED> " + app.get('port'));
});
    \end{lstlisting}
	\end{subfigure}\hfill
	\begin{subfigure}{.495\columnwidth}
		\begin{lstlisting}[language=jsappdx, title={(b) Output: non-vulnerable full file}, captionpos=t, numberstyle=\linecolors{white}{0}{0}]		
var requirejs = require('requirejs');
var config = requirejs('./config');

// ... <REDACTED>
    
var express = require('express'),
(*\HL{mygreen}*)    https = require('https'),
    path = require('path');

// ... <REDACTED>
    
var rootDir = path.join(__dirname, '..');

var app = express();

app.configure(function(){
  app.set('port', process.env.PORT || 3000);
  app.use(express.logger('dev'));
  // ... <REDACTED>
});

app.configure('development', function(){
  app.use('/app', 
    express.static(path.join(rootDir, 'app')));
  // ... <REDACTED>
});

(*\HL{mygreen}*)var server = https.createServer(app)
.listen(app.get('port'), function(){
  console.log("<REDACTED> " + app.get('port'));
});
	  \end{lstlisting}
	\end{subfigure}

	%
	%
	%
	%
	\caption{\tool{} fixes a \textit{HttpToHttps} vulnerability requiring multiple changes in different locations of the file.
	}
\end{figure*}

\clearpage{}
\section{Evaluation of Model Size and Architechture with CodeReduce}\label{sec:appendix_size_arch}
\begin{table*}[h]
	\caption{Effects of the model size and architecture for \JsReducedData{} data, w.r.t. $\PassAtKMetric{}$ for $k=1$.}
	\centering
	\scriptsize
	\begin{tabular}[c]{@{\quad}l*{3}{>{\centering\arraybackslash}p{\widthof{XXXXXXXXXXXXXXXXXX}}}@{\quad}}
		\toprule
		\mulro[.6]{2}{Issue Category} & \multicolumn{3}{c}{\PassAtKMetric{} ($\%$), $k=1$}                                                            \\
		\cmidrule(l{1pt}r{1pt}){2-4}
		                              & \textsc{StarCoderBase-1B}                          & \textsc{StarCoderBase-7B} & \textsc{Mistral-7B-Instruct} \\
		\midrule
		\AST                          & $59.00$                                            & $\underline{\bm{70.05}}$  & $63.78$                      \\
		\cmidrule(l{1pt}r{1pt}){1-4}
		\Local                        & $72.02$                                            & $\underline{\bm{82.75}}$  & $70.09$                      \\
		\cmidrule(l{1pt}r{1pt}){1-4}
		\FileWide                     & $54.28$                                            & $\underline{\bm{77.38}}$  & $51.48$                      \\
		\cmidrule(l{1pt}r{1pt}){1-4}
		\SecurityLocal                & $67.06$                                            & $\underline{\bm{74.68}}$  & $56.33$                      \\
		\cmidrule(l{1pt}r{1pt}){1-4}
		\SecurityFlow                 & $39.64$                                            & $\underline{\bm{49.15}}$  & $27.47$                      \\

		\bottomrule
	\end{tabular}
	\\[.3cm]
	\label{table:passk-long-vs-base}
\end{table*}

\section{Evaluation of Different Models with CodeReduce}\label{sec:appendix_codereduce_models}

\begin{table*}[!htbp]
	\centering
	\scriptsize
	\renewcommand{\arraystretch}{1.2}
	\caption{Evaluation of \PassAtKMetric{} and \ExactMatchAtKMetric{} metrics for models that use \codereduce{} (marked as $^{\bm{\ddag}}$)}
	\begin{tabular}[c]{@{\quad}ll*{6}{>{\centering\arraybackslash}p{\widthof{\;$k=5$\;}}}@{\quad}}
		\toprule
		\mulro[.6]{2}{Issue Category} & \mulro[.6]{2}{Model}
		                              & \multicolumn{3}{c}{\PassAtKMetric{} ($\%$)}                                                        & \multicolumn{3}{c}{\ExactMatchAtKMetric{} ($\%$)}                                                   \\
		\cmidrule(l{1pt}r{2pt}){3-5} \cmidrule(l{1pt}r{2pt}){6-8}
		                              &                                                                                                    & $k = 1$                                           & $k=3$   & $k = 5$ & $k = 1$ & $k=3$   & $k = 5$ \\
		\midrule
		\mulro{5}{\AST}
		                              & \rlap{\textsc{Mixtral-8x7B}}\phantom{\textsc{Mixtral-8x7B}}\textsc{-\JsReducedData}$^{\bm{\ddag}}$

		                              & $81.01$                                                                                            & $88.96$                                           & $89.16$ & $39.29$ & $52.25$ & $53.5$            \\

		                              & \rlap{\textsc{StarCoder-7B}}\phantom{\textsc{Starcoder-7B}}\textsc{-\JsReducedData}$^{\bm{\ddag}}$

		                              & $70.05$                                                                                            & $80.84$                                           & $82.31$ & $41.19$ & $48.15$ & $48.56$           \\

		                              & \textsc{StarCoder-3B-\JsReducedData{}}$^{\bm{\ddag}}$

		                              & $68.2$                                                                                             & $78.86$                                           & $81.84$ & $40.51$ & $43.79$ & $43.79$           \\

		                              & \textsc{StableCode-3B-\JsReducedData{}}$^{\bm{\ddag}}$

		                              & $67.21$                                                                                            & $79.77$                                           & $80.6$  & $37.69$ & $46.51$ & $47.02$           \\

		                              & \textsc{DeepSeekCoder-1.3B-\JsReducedData{}}$^{\bm{\ddag}}$
		                              & $57.39$                                                                                            & $75.41$                                           & $79.27$ & $30.75$ & $39.67$ & $45.48$           \\

		\midrule
		\mulro{5}{\Local}
		                              & \rlap{\textsc{Mixtral-8x7B}}\phantom{\textsc{Mixtral-8x7B}}\textsc{-\JsReducedData}$^{\bm{\ddag}}$

		                              & $69.77$                                                                                            & $90.19$                                           & $92.71$ & $31.37$ & $41.77$ & $44.11$           \\

		                              & \rlap{\textsc{StarCoder-7B}}\phantom{\textsc{StarCoder-7B}}\textsc{-\JsReducedData}$^{\bm{\ddag}}$

		                              & $82.75$                                                                                            & $89.72$                                           & $90.19$ & $34.23$ & $42.02$ & $42.73$           \\

		                              & \textsc{StarCoder-3B-\JsReducedData{}}$^{\bm{\ddag}}$
		                              & $81.87$                                                                                            & $85.82$                                           & $87.63$ & $33.99$ & $40.79$ & $44.02$           \\

		                              & \textsc{StableCode-3B-\JsReducedData{}}$^{\bm{\ddag}}$

		                              & $77.51$                                                                                            & $83.39$                                           & $83.93$ & $32.18$ & $39.72$ & $40.72$           \\

		                              & \textsc{DeepSeekCoder-1.3B-\JsReducedData{}}$^{\bm{\ddag}}$
		                              & $72.72$                                                                                            & $82.24$                                           & $85.05$ & $33.98$ & $42.97$ & $45.4$            \\

		\midrule
		\mulro{5}{\FileWide}
		                              & \rlap{\textsc{Mixtral-8x7B}}\phantom{\textsc{Mixtral-8x7B}}\textsc{-\JsReducedData}$^{\bm{\ddag}}$

		                              & $50.37$                                                                                            & $67.61$                                           & $85.66$ & $41.66$ & $47.91$ & $48.95$           \\


		                              & \rlap{\textsc{StarCoder-7B}}\phantom{\textsc{StarCoder-7B}}\textsc{-\JsReducedData}$^{\bm{\ddag}}$
		                              & $77.38$                                                                                            & $80.16$                                           & $85.02$ & $44.59$ & $54.5$  & $54.5$            \\

		                              & \textsc{StarCoder-3B-\JsReducedData{}}$^{\bm{\ddag}}$

		                              & $61.88$                                                                                            & $81.9$                                            & $82.59$ & $38.69$ & $57.91$ & $73.11$           \\

		                              & \textsc{StableCode-3B-\JsReducedData{}}$^{\bm{\ddag}}$

		                              & $69.74$                                                                                            & $76.34$                                           & $76.34$ & $38.4$  & $40.98$ & $40.98$           \\

		                              & \textsc{DeepSeekCoder-1.3B-\JsReducedData{}}$^{\bm{\ddag}}$
		                              & $64.98$                                                                                            & $89.29$                                           & $94.89$ & $34.53$ & $62.91$ & $65.27$           \\

		\midrule
		\mulro{5}{\SecurityLocal}
		                              & \rlap{\textsc{Mixtral-8x7B}}\phantom{\textsc{Mixtral-8x7B}}\textsc{-\JsReducedData}$^{\bm{\ddag}}$

		                              & $63.58$                                                                                            & $92.2$                                            & $94.41$ & $15.77$ & $23.46$ & $25.99$           \\


		                              & \rlap{\textsc{StarCoder-7B}}\phantom{\textsc{StarCoder-7B}}\textsc{-\JsReducedData}$^{\bm{\ddag}}$

		                              & $74.68$                                                                                            & $89.08$                                           & $91.76$ & $20.18$ & $29.53$ & $32.23$           \\

		                              & \textsc{StarCoder-3B-\JsReducedData{}}$^{\bm{\ddag}}$

		                              & $68.04$                                                                                            & $88.51$                                           & $91.41$ & $17.86$ & $25.6$  & $27.43$           \\

		                              & \textsc{StableCode-3B-\JsReducedData{}}$^{\bm{\ddag}}$

		                              & $70.94$                                                                                            & $84.23$                                           & $86.49$ & $15.31$ & $18.71$ & $23.36$           \\

		                              & \textsc{DeepSeekCoder-1.3B-\JsReducedData{}}$^{\bm{\ddag}}$
		                              & $61.69$                                                                                            & $84.18$                                           & $88.66$ & $15.99$ & $26.84$ & $28.82$           \\

		\midrule
		\mulro{5}{\SecurityFlow}

		                              & \rlap{\textsc{Mixtral-8x7B}}\phantom{\textsc{Mixtral-8x7B}}\textsc{-\JsReducedData}$^{\bm{\ddag}}$

		                              & $41.93$                                                                                            & $72.97$                                           & $81.62$ & $9.26$  & $17.9$  & $20.27$           \\


		                              & \rlap{\textsc{StarCoder-7B}}\phantom{\textsc{StarCoder-7B}}\textsc{-\JsReducedData}$^{\bm{\ddag}}$

		                              & $49.15$                                                                                            & $65.9$                                            & $68.53$ & $10.63$ & $13.49$ & $15.7$            \\

		                              & \textsc{StarCoder-3B-\JsReducedData{}}$^{\bm{\ddag}}$

		                              & $45.65$                                                                                            & $58.82$                                           & $65.99$ & $9.72$  & $16.17$ & $21.07$           \\

		                              & \textsc{StableCode-3B-\JsReducedData{}}$^{\bm{\ddag}}$

		                              & $41.28$                                                                                            & $55.32$                                           & $58.58$ & $9.86$  & $15.37$ & $15.78$           \\

		                              & \textsc{DeepSeekCoder-1.3B-\JsReducedData{}}$^{\bm{\ddag}}$
		                              & $33.26$                                                                                            & $59.61$                                           & $66.88$ & $15.22$ & $19.27$ & $19.77$           \\

		\bottomrule
	\end{tabular}
	\\[.2cm]
	\label{table:results-reduced-vs-window}
	\vspace{-.75cm}
\end{table*}

\end{document}